\documentclass[sigconf]{acmart}

\AtBeginDocument{%
  }

\copyrightyear{2026}
\acmYear{2026}
\setcopyright{cc}
\setcctype{by}
\acmConference[CHI '26]{Proceedings of the 2026 CHI Conference on Human Factors in Computing Systems}{April 13--17, 2026}{Barcelona, Spain}
\acmBooktitle{Proceedings of the 2026 CHI Conference on Human Factors in Computing Systems (CHI '26), April 13--17, 2026, Barcelona, Spain}
\acmPrice{}
\acmDOI{10.1145/3772318.3791281}
\acmISBN{979-8-4007-2278-3/2026/04}

\usepackage{color, colortbl}
\usepackage{subcaption, multirow}
\usepackage{soul} 
\usepackage{graphicx}

\definecolor{light-gray}{gray}{0.95}
\definecolor{airforceblue}{rgb}{0.36, 0.54, 0.66}
\definecolor{majorelleblue}{rgb}{0.38, 0.31, 0.86}
\definecolor{oceanboatblue}{rgb}{0.0, 0.47, 0.75}
\definecolor{st}{HTML}{538E64}
\definecolor{c}{HTML}{56B4E9}
\definecolor{se}{HTML}{F5A42B}
\definecolor{oc}{HTML}{CC79A7}
\usepackage{enumitem}

\begin{document}
\title{Value Alignment of Social Media Ranking Algorithms}

\author{Farnaz Jahanbakhsh}
\authornote{Equal contribution}
\affiliation{%
  \institution{University of Michigan}
  \city{Ann Arbor, MI}
  \country{USA}
}
\email{farnaz@umich.edu}

\author{Dora Zhao}
\affiliation{%
  \institution{Stanford University}
  \city{Stanford, CA}
  \country{USA}
}
\email{dorothyz@stanford.edu}
\authornotemark[1] 

\author{Tiziano Piccardi}
\affiliation{%
  \institution{Johns Hopkins University}
  \city{Baltimore, MD}
  \country{USA}
}

\author{Zachary Robertson}
\affiliation{%
  \institution{Stanford University}
  \city{Stanford, CA}
  \country{USA}
}

\author{Ziv Epstein}
\affiliation{%
  \institution{Massachusetts Institute of Technology}
  \city{Cambridge, MA}
  \country{USA}
}

\author{Sanmi Koyejo}
\affiliation{%
  \institution{Stanford University}
  \city{Stanford, CA}
  \country{USA}
}

\author{Michael S. Bernstein}
\affiliation{%
  \institution{Stanford University}
  \city{Stanford, CA}
  \country{USA}
}

\renewcommand{\shortauthors}{Jahanbakhsh and Zhao et al.}

\begin{abstract}
While social media feed rankings are primarily driven by engagement signals rather than any explicit value system, the resulting algorithmic feeds are not value-neutral: engagement may prioritize specific individualistic values. This paper presents an approach for social media feed value alignment.
We adopt Schwartz’s theory of Basic Human Values --- a broad set of human values that articulates complementary and opposing values forming the building blocks of many cultures --- and we implement an algorithmic approach that models and then ranks feeds by expressions of Schwartz's values in social media posts. Our approach enables controls where users can express weights on their desired values, combining these weights and post value expressions into a ranking that respects users' articulated trade-offs. Through controlled experiments ($N=141$ and $N=250$), we demonstrate that users can use these controls to architect feeds reflecting their desired values. Across users, value-ranked feeds align with personal values, diverging substantially from existing engagement-driven feeds.

\end{abstract}

\ccsdesc[500]{Human-centered computing~Collaborative and social computing systems and tools}

\keywords{social media, algorithms, feed curation}

\begin{teaserfigure}
\begin{minipage}{\textwidth}
  \includegraphics[width=\textwidth]{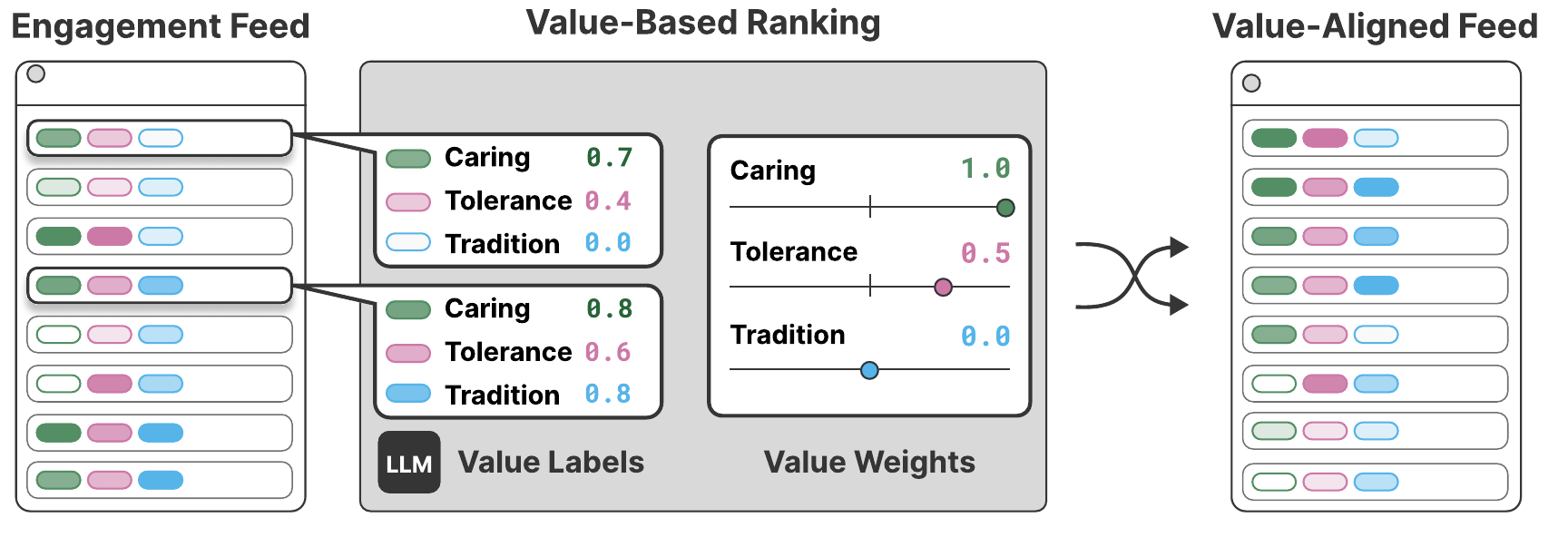}
\end{minipage}
    \caption{We describe a method for value alignment to rank social media feeds using basic human values. For a given inventory of social media content, we classify value expressions using a large language model. We combine these value labels with a set of weights reflecting how much a given value ought to be amplified or dampened in order to score social media posts. The weights can be learned implicitly or directly specified by the users. Finally, the scores are then used to rank content within a feed.}
    \label{fig:teaser}
    \Description[Diagram illustrating the proposed value-aligned feed ranking method.]{
    Diagram illustrating the proposed value-aligned feed ranking method. Social media posts are classified by a large language model into Schwartz’s values, combined with user-specified weights, and ranked accordingly. The figure shows an example where weights for “Caring,” “Tolerance,” and “Tradition” are specified, demonstrating how these weights alter the ranking.}
\end{teaserfigure}

\maketitle
\section{Introduction}
Value alignment has been at the center of public and academic debate in recent years, and especially regarding generative AI~\cite{khamassi2024strong, klingefjord2024human,huang2025values}. But generative models are not the only systems shaping our informational environment. Social media feeds too, are products of algorithmic systems that curate what we see and mold how we experience the world~\cite{brady2023overperception, pennycook2018prior}. 
And like with generative AI~\cite{jakesch2023co, bai2025llm}, the behavior of these models shapes our attitudes, beliefs, and behavior, so the values they encode --- empathy or self-interest, integrity or convenience --- can influence both individual and societal outcomes~\cite{piccardi2024reranking}.
Yet, feed algorithms continue to operate with little consideration for the values embedded in the content they amplify~\cite{bernstein2023embedding}.
 
Instead, feed algorithms primarily optimize for predicted engagement signals such as clicks, shares, and likes~\cite{cunningham2024we,narayanan2023,eckles2022algorithmic,klug2021trick}. The product of this ``value-free'' optimization goal is far from value-neutral; by prioritizing content that excites, surprises, or provokes, the algorithms curate feeds that disproportionately emphasize specific, short-term individualistic values such as enjoyment and novelty. Both critical (e.g.,~\cite{munn2020angry}) and empirical (e.g.,~\cite{allcott2020welfare,feng2024mapping}) work has documented the detrimental effects of our feed algorithms as positioned today. 

How might we --- as platform users, as platform operators, or as a society --- consider value alignment of these feed algorithms? How can we be intentional about which values are amplified and dampened through  algorithmic feed ranking? Could we envision a system where other types of values are prioritized --- perhaps well-being or camaraderie, which might better align with an individual's longer-term aspirations or benefit society more broadly?

To value align feed algorithms, in this paper, we develop an approach that operationalizes a structured value system --- one that is grounded in the empirical behavioral literature and accounts for trade-offs between values --- into social media ranking algorithms. Specifically, we adopt Schwartz’s theory of Basic Human Values~\cite{schwartz2012refining}, which has been widely used and replicated in Cultural Psychology. The modern Schwartz theory articulates 19 conceptually distinct values, each of which is placed into a circumplex model to articulate values that are complementary or in tension with each other. For example, Achievement captures an individualistic focus, whereas Caring represents an opposite and complementary societal focus. The Cultural Psychology literature argues that these values are the basic motivations that drive behavior universally across cultures~\cite{schwartz2012overview}, and can be thought of as the building blocks of other, more specific, values.

We contribute an algorithm for classifying expressions of Schwartz's Basic Human Values in social media posts, then use these models to rerank social media feeds (Fig.~\ref{fig:teaser}). We draw on the insight that the value definitions defined by social scientists and humanities scholars are precise enough constructs~\cite{schmitt1993measurement} for large language models (LLMs) to annotate as accurately as experts do~\cite{jia2024embedding,halterman2024codebook,tai2024examination,dai2023llm,ziems2024can}. Leveraging this insight, we develop a technical approach to classify social media posts using Schwartz's values, then integrate those classifications with users' value priorities in order to rerank their feed to be aligned with their desired values. 

But even if we technically rerank feeds to reflect different values, we need some assurance that users' value specifications are preserved in a meaningful way rather than being collapsed into noise. When multiple values interact, and often stand in tension, recognizability of the values in the resulting feed becomes a necessary condition for establishing that the feed meaningfully reflects those specifications. If users can perceive a coherent relationship between the values they articulate and the feed they receive, then the pipeline from elicitation to ranking is at least preserving a meaningful input–output mapping.

We demonstrate through two controlled experiments that users can distinguish between value-aligned and engagement-aligned rankings of their own social media feeds, and that this recognizability weakens as users apply more complicated combinations of values. In both experiments, we rank users' own Twitter/X feeds by the values that they deem important, and present them along the default engagement feed in a blinded selection task.

In the first experiment, we derive users' self-reported value priorities from a value questionnaire that they fill out. In the second, we offer users value controls that they can directly manipulate and whose effects on their feed they can observe in real time. We take the configured values and apply them to users' own Twitter feeds, then task users with identifying which is the value-aligned feed and which is the original engagement-aligned feed. Finally, through analyzing 188 re-ranked feeds, we find that users often configure controls that align with their personal values, leading to value-aligned feeds that diverge substantially from engagement-driven feeds (mean Kendall’s $\tau$ of $0.06$).

In sum, this work contributes:
\begin{enumerate}
    \item \textbf{A framework for value-aligning feed ranking algorithms}: We introduce the idea of ranking social media feeds by a value system that is both broadly encompassing and grounded in trade-offs between values. We present a method for incorporating multiple values into content ranking and instantiate it using the \emph{Schwartz Basic Theory of Human Values} to annotate expressions of values in social media posts.
    \item \textbf{Two online experiments validating our approach}: \newline
    Through controlled studies, we demonstrate that our approach yields feed rankings that recognizably reflect participants' articulated values.
    \item \textbf{Empirical insights into value selections}: We analyze what mechanisms drive users' value selections and how these selections impact feed rankings.
\end{enumerate}
Ultimately, this work advocates that social media feeds need neither disregard values nor selectively prioritize only a few, but can consider a broad space of value tradeoffs and empower users, platforms, and society to better navigate them.\footnote{Code and project website available at \url{https://stanfordhci.github.io/FeedValueAlignment/}.}

\section{Related Work}
\subsection{Incorporating human values in ranking algorithms}
At the moment, social media platforms primarily utilize engagement signals, such as number of clicks, time spent on an item, or number of comments, to inform feed ranking~\cite{backstrom2016serving,narayanan2023,bernstein2023embedding,cunningham2024we,milli2025engagement}. Compared to chronological feeds, engagement-based feeds increase user retention~\cite{cunningham2024we, guess2023social, bandy2023exposure}. However, focusing solely on engagement signals can lead to detrimental social outcomes, including the marginalization of perspectives~\cite{harris2023honestly,delmonaco2024you}, proliferation of misinformation~\cite{valenzuela2019paradox, jahanbakhsh2021exploring, hao2021facebook}, political polarization~\cite{rathje2021out,kubin2021role}, and promotion of extremist content~\cite{o2015down}. It is not only what content users see, but also the ranking of the information can impact users' opinion formation and even influence downstream actions, such as swaying voting preferences~\cite{epstein2015search,draws2021not,chan2025ranking}. To address these issues, recent work has argued that the signals we incorporate into ranking algorithms should be expanded to also take into account broader societal values~\cite{bernstein2023embedding}. While this commitment may lead to tradeoffs with engagement, early evidence did not find measurable engagement tradeoffs when downranking content that exhibited anti-democratic attitudes~\cite{jia2024embedding}.

This principle --- that we ought to incorporate values into technical design --- is not new to human-computer interaction. In the 1990s, \citet{friedman2013value} introduced Value-Sensitive Design, which proposed that designers must consider human values, or ``what a person or group of people consider important in life'', when creating technical systems. \citet{zhu2018value} adapted this framework and proposed ``Value-Sensitive Algorithm Design,'' which centers on engaging relevant stakeholders when creating machine learning systems. In this vein, subsequent work has proposed methods to facilitate deliberation with community members on societal concerns regarding machine learning~\cite{shen2021value, shen2022model}. In addition to interventions implemented during the design process, human values can also be used to steer model outputs. For example, techniques, such as Reinforcement Learning through Human Feedback (RLHF)~\cite{bai2022training} or Constitutional AI~\cite{bai2022constitutional}, are used to attempt to align LLM's behavior towards ``desirable'' values. Nonetheless, these models are still predisposed to produce outputs promoting a narrow swath of values, such as those that are more Western-centric~\cite{ryan2024unintended, durmus2023towards} or concordant with left-leaning beliefs~\cite{santurkar2023whose}. 

The challenging questions are: (1)~\emph{which} social values should be embedded in algorithms, and (2)~\emph{how} do we do so? In the social media setting, many prior works have focused on political aspects of this problem, motivated by the pressing concerns about these technologies' impact on democracy~\cite{jia2024embedding,bluefeedredfeed,babaei2018purple,nelimarkka2019re, munson2010presenting, munson2013encouraging, guess2023reshares, nyhan2023like,milli2025engagement}. There have also been concerted efforts to reduce the amount of misinformation on feeds~\cite{jahanbakhsh2021exploring,bode2018see,karduni2019vulnerable,vraga2017using}. However, focusing only on misinformation or political content covers only a small fraction of what is shown on social media. Perhaps closest to our work is Alexandria~\cite{kolluri2025alexandria}, a library of values spanning six value systems (e.g., Rokeach and Maslow). While Alexandria offers broad coverage, it does not encode relationships of tension or complementarity among values; this limitation is important as algorithms must inevitably prioritize some values at the expense of others. In contrast, Schwartz's value system provides a theoretical structure for reasoning over trade-offs between values. Furthermore, our work validates that our method using direct user elicitation and LLM value labels produces recognizably value-aligned feed and additionally provides empirical insights into how this value alignment reshapes users' feeds. Finally, our studies reveal that maintaining recognizability becomes difficult when users adjust multiple values simultaneously, though this decline plateaus rather than continuing to degrade. This finding has concrete implications for designing value elicitation mechanisms. Outside of Alexandria, other efforts to advocate for a broader set of values have remained at the theoretical level~\cite{bernstein2023embedding,stray2024building,milli2021optimizing,rieger2024responsible}. This leaves questions about how we might integrate a \emph{design space of values} --- one with trade-offs --- into algorithms in practice. 

This work presents a method for ranking social media feeds using a value system built around trade-offs between values (Schwartz's Basic Human Values) and by empirically testing whether feeds optimized for multiple values at once still produce outputs that users can recognize as aligned with their selected values or whether such multi-objective optimization might otherwise collapse into unrecognizable blends. To define the values present in social media content, we draw from Schwartz's theory of Basic Human Values, which articulates a encompassing set of values that underlie the attitudes and behaviors of individuals. We can use large language models to label these constructs at scale and then incorporate them into the feed-ranking algorithm. We demonstrate that our proposed method is robust and flexible enough to be successfully applied to users' own feeds.


\subsection{Providing end-user control on social media}

The approach that we detail in this paper can be executed at the central platform level (e.g., Facebook, Twitter, Instagram, and TikTok), in a decentralized way by individual communities (e.g., if Mastodon added ranked feeds), or by end users themselves (e.g., Bluesky). As our approach allows end users themselves to specify values, in this section, we review how and why increased community and end user control of feed algorithms can be beneficial.

In an age when much of the information consumption happens on social media by choice or accident, the platforms as the authorities that govern the flow of information wield significant influence over public discourse, shaping narratives and potentially narrowing the diversity of viewpoints~\cite{draws2021assessing,bozdag2015breaking,kitchens2020understanding}. This concentration of power has raised concerns, both in the academic sphere and among the general public, about bias, censorship, or the ad-hoc development and inconsistent enforcement of rules and policies~\cite{saltz2021encounters, diaz2021double, schneider2024governable}. These concerns are grounded in accounts of platforms removing content that arguably posed little to no harm~\cite{facebookCensorshipActivistsOfColor, FacebookPragerU, instagramFailingProtesters}, disproportionately policing speech from minority groups~\cite{facebookCensorshipActivistsOfColor, facebookCensorshipNWord, sanchez2024observers}, selectively intervening in certain high-profile cases by bending or outright breaking community standards~\cite{Facebook2022UkraineRussia, FacebookHateSpeechPolicy}, and pursuing inaction in others with grave ramifications~\cite{MyanmarFacebook, Facebook2022UkraineRussia}, among other troubling practices.

Even if we do not view platforms as unfit governors~\cite{klonick2017new} of speech, we must still seriously weigh the benefits of granting end users more control over their feed experiences. Centralized decisions about what content to show or suppress can fail to account for the diverse needs and desires of each user or community. For example, while platforms typically down-rank misinformation, prior work has shown that some users prefer to avoid unreliable content entirely, while others want to see it to understand what misinformation their friends and family see in order to talk to them about it~\cite{jahanbakhsh2022leveraging, malhotra2020covid19, rader2015understanding}. As another example, minority groups, who use social media to call out racism and recount the language used against them, report being found in violation of centralized guidelines, as their posts are misconstrued as hate speech~\cite{facebookCensorshipNWord, lee2024people}.

The centralized control of information on platforms is primarily executed through content curation and moderation, much of which is handled algorithmically. In the absence of end-user controls explicitly built into the process, users resort to finding ways around the algorithm, albeit with uncertain efficacy, by developing ``folk theories'' that rationalize how their inputs translate into outputs~\cite{eslami2015always}. For instance, to avoid seeing content they do not like, they might \textit{``like other things such that these show up less''}~\cite{jahanbakhsh2022leveraging, eslami2016first}.

With the growing awareness about the shortcomings of one-size-fits-all approaches and the recognition that users do desire control, there have been endeavors to give users more agency over their information space. Notable examples in the industry include Mastodon, a decentralized social network where users can join servers with varying curation and moderation policies, and Bluesky, which allows users to subscribe to different feed algorithms or moderation modules~\cite{bluesky}. Tools such as Skyfeed\footnote{https://skyfeed.app/} and Graze\footnote{https://www.graze.social/} utilize Bluesky's API, offering users the ability to author custom feeds. A body of academic work has also concerned itself with pluralist models of governance that range from delegating the power of decision-making to community juries~\cite{fan2020digital, micek2024examining}, to empowering the individual user~\cite{jhaver2023users,burrell2019users}. These have been instantiated in the form of filters or extensions on top of existing platforms~\cite{bhargava2019gobo, jahanbakhsh2022our, jahanbakhsh2024browser, jhaver2023personalizing} as well as separate social media platforms~\cite{jahanbakhsh2022leveraging}. Another line of work has proposed new methods for personalizing the feed experience using AI tools~\cite{feng2024mapping, jahanbakhsh2023exploring,wang2025end}.

Building on this literature, we design an interface that allow users to specify their value preferences and directly reconfigure their feeds. While prior work focused on empowering interventions in more targeted use cases, such as reducing misinformation~\cite{jahanbakhsh2022leveraging} and combating harassment~\cite{jhaver2023personalizing}, or only provided a limited set of filters~\cite{bhargava2019gobo}, we present a more comprehensive set of controls. By focusing on the \emph{basic} values present in the feed, and incorporating those into the algorithm, our controls can be applied to the broad range of content that is available on social media. 

There remains debate on how control and power should be split between platforms, communities, and end users. In this paper, we do not pretend to resolve this debate. Our approach most directly offers tools and levers for end users, but also opportunities for platforms. The Discussion section will reflect in more detail about the risks of centralized platforms or governments using this approach to force values on their population, and the mitigations to that risk that we would advocate for.

\section{Value Aligning Social Media Feeds with Schwartz's Basic Values}
\label{section:identification-ranking-process}
To enable our social media feeds to become responsive to values and their tradeoffs, we require a complete value system: a set of values that anchor the basic tensions at stake. Of course, there are many possible values (i.e., \citet{sorensen2024value} articulated over two hundred thousand values, rights, and duties). Instead of seeking an articulation of every single value, we instead advocate for structural coverage in the sense that the set of values encompasses the main axes of differentiation in the design space, and that most specific values can be thought of as specializations of those in our set. 

In our work, we use Schwartz's theory of basic values, which provides a set of universally recognized concepts, for operationalizing these constructs in social media content. Schwartz's system has been applied not only for looking at individuals' values but also for analyzing values in text, including news articles and arguments~\cite{bardi2008new,kiesel2022identifying,borenstein2024investigating}. 

According to Schwartz, values transcend specific situations and serve as standards or criteria for what is considered good or bad. Trade-offs among these broad motivational principles guide attitudes and behaviors across a wide range of contexts. For example, values such as obedience or honesty may be relevant whether one is interacting with friends, colleagues, or strangers~\cite{schwartz2012overview, schwartz1992universals, schwartz2013value}. Schwartz further shows that individuals' general value priorities, elicited in a domain-independent way, predict patterns of behavior across diverse contexts, from interpersonal cooperation to voting behavior to intergroup contact~\cite{schwartz2013value}.
This trans-situational quality distinguishes values from norms or attitudes, which tend to be context-specific~\cite{schwartz2012overview}. Because Schwartz's values operate at this general level and have been extensively validated across diverse contexts, they can be elicited in a domain-general form and then operationalized in specific contexts such as social media consumption.

While Schwartz's theory is not the only lens through which values can be investigated, it does provide a comprehensive value system that has been extensively validated across different populations~\cite{schwartz1990toward,schwartz2014values,schwartz2017value}.  Of course, other complete value sets may also satisfy these criteria (e.g., Moral Foundations Theory~\cite{graham2013moral}). Our method aims to generalize to any value system that satisfies these criteria.

Given the value system, our goal is to operationalize that complete set of values into algorithms, elicit how users prioritize the values, and incorporate these priorities into a feed ranking model. To scale the coding process, we use large language models to classify value labels. Finally, given user input on value priorities, we are able to re-order feeds to better reflect these preferences.

\subsection{Measuring value expressions in social media posts}

\begin{figure}
    \centering
    \includegraphics[width=0.7\linewidth]{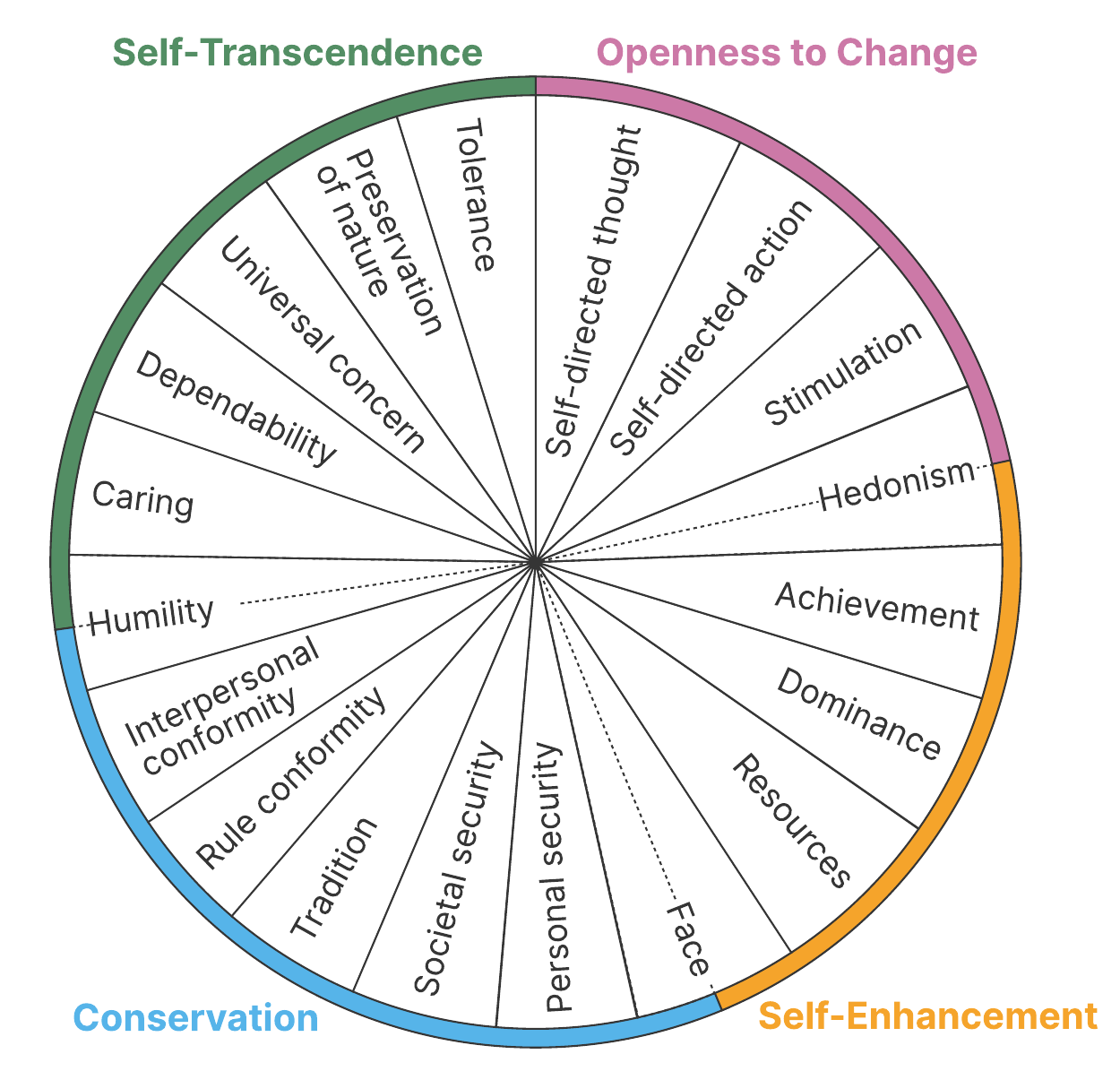}
    \caption{Visualization of Schwartz's 19 Basic Human Values. Values close to each other in the circle tend to share similar motivations whereas opposing values often are in tension with each other. The 19 values are also organized into four broader groups: self-transcendence, conservation, self-enhancement, and openness to change.}
    
    \label{fig:circ}
    \Description[Circular diagram of Schwartz’s 19 Basic Human Values.]{
    Circular diagram of Schwartz’s 19 Basic Human Values. Values are organized on a circumplex, where adjacent values share similar motivations, and opposing values are in tension. The four broader groups—self-transcendence, conservation, self-enhancement, and openness to change—are highlighted
    }
\end{figure}

\begin{table*}[]
    \footnotesize
    \centering
    \begin{tabular}{p{1in}p{1in}p{3.4in}}
    \toprule
    \small{\textbf{Schwartz's Value}} &  \small{\textbf{User-Facing Title in Our System}} & \small{\textbf{Definition}} \\
    \midrule
    \textit{Openness to Change} & \textit{Openness to Change}\\
    \midrule
    Self-directed thoughts & Independent thoughts & The freedom to cultivate one's own ideas and abilities\\
    Self-directed actions & Independent actions & The freedom to determine one's own actions\\
    Stimulation & Novelty & Excitement, stimulation, and change\\ 
    Hedonism & Pleasure & Hedonism\\ 
    \midrule
    \textit{Self-Enhancement} & \textit{Personal Growth}\\ 
    \midrule
    Achievement & Achievement & Success according to social standards\\
    Dominance & Power & Influence and the right to command\\ 
    Resources & Wealth & Control of material and social resources\\ 
    Face & Reputation & Security and power through maintaining one's public image and avoiding humiliation\\
    \midrule
    \textit{Conservation} & \textit{Stability}\\ 
    \midrule
    Personal security & Personal security & Safety in one's immediate environment\\
    Societal security & Societal security & Safety and stability in the wider society\\ 
    Tradition & Tradition & Maintaining and preserving cultural, family, or religious traditions\\
    Rule conformity & Lawfulness & Compliance with rules, laws, and formal obligations\\ 
    Interpersonal conformity & Respect & Avoiding upsetting or harming other people\\ 
    Humility & Humility & Being humble\\
    \midrule
    \textit{Self-Transcendence} & \textit{Selflessness}\\
    \midrule
    Caring & Caring & Devotion to those they care about\\ 
    Dependability & Responsibility & Being responsible and having loyalty to others\\ 
    Universal concern & Equality & Commitment to equality, justice, and protection for all people\\ 
    Preservation of nature & Connection to nature & Preservation of the natural environment\\
    Tolerance & Tolerance & Acceptance and understanding of those different from oneself\\
    \bottomrule
    \end{tabular}
    \caption{Overview of the 19 values in Schwartz's refined theory of basic human values. The table contains the value name that Schwartz used in the original theory, the simplified name for the value that participants were shown in Study 2, and a definition of the value that was also provided to participants.}
    \label{tab:value_names}
\end{table*}

We start by developing a method for identifying values in social media content. Our goal is to adapt Schwartz's theory of Basic Human Values so that we can identify and operationalize expressions of values in social media posts. Schwartz's theory is designed so that the values are recognized across societies as guiding the decisions and behaviors of individuals. In particular, psychologists have leveraged this theory for understanding what values different cultures prioritize~\cite{schwartz2004mapping,schwartz2006theory}. 

\subsubsection{Classifying values} Because the literature has articulated several versions of Schwartz's theory, we adapt Schwartz's most recent refined theory of Basic Human Values, which consists of 19 distinct values. As shown in Fig.~\ref{fig:circ}, these values are organized on a circumplex, reflecting that the ones adjacent to each other share similar underlying motivations, while the ones on opposite sides of the circle have conflicting motivations. Schwartz further synthesizes these fine-grained values into higher-level clusters of Self-Transcendence, Openness to Change, Conservation, and Self-Enhancement. Definitions for each value are listed in Table~\ref{tab:value_names}.

To identify values in tweets, we consider both the expression of the value and the magnitude. We score each value's expression on each post as an integer value. A score of 0 indicates that the value does not exist or is not supported in the tweet (i.e., the tweet contains content that contradicts a value). If a tweet does contain content that supports the value, the scores range from 1 (i.e., the value is slightly reflected in the content) to 6 (i.e., the value is strongly reflected). We do not claim that this rating scale is the only possible one; however, we advocate for an approach that can model whether values are present or not as well as an ordinal or continuous strength scale.

Next, to allow users to re-rank their feeds, we need an automated way to code value expression scores~\cite{piccardi2024reranking}. As prior work has demonstrated~\cite{jia2024embedding}, LLMs can apply constructs from social sciences for coding social media content. In our work, we use few-shot prompting with \texttt{GPT-4o} to code for Schwartz's values in tweets. \texttt{GPT-4o} and similar multi-modal models (e.g., \texttt{GPT-4o-mini}, \texttt{Gemini}, and \texttt{Llama}) are an attractive option for this task because their training has exposed them to a broad range of cultural, political, and topic themes, and their ability to take in images as well as text fits well with the social media context. As input, we give the model the text of the tweet, accompanying images, and embedded links along with their descriptions.
In addition to two examples that were manually coded by the researchers, we provide definitions of the 19 values taken from Schwartz's theory, and the tweet itself, as input to the model. Since Schwartz argues that values are in fact not independent from each other and share the same (or conflicting) underlying motivations, we included all 19 values in one prompt rather than building individual classifiers. As output, the model returns ratings for all 19 values in JSON format. See the Appendix for prompt details and examples of coded tweets. 

\subsubsection{Validating classifier performance} 
To validate the performance of the LLM classifier, we compare the generated labels with a dataset of 4,562 Twitter/X posts from \citet{epstein2025measuring}, in which each post is labeled by $\geq 4$ human annotators. For each post, we report LLM-Consensus MAE, or the mean absolute error (MAE) between the LLM-generated label and the consensus label (e.g., mean human rating). We then compare this error to the mean disagreement that an individual human annotator has with the consensus label (Human-Consensus MAE). To formalize this, for each of the $n$ annotators who labeled a value $v$ on a given post, we compute the MAE between the label they assigned to the value and the mean labels of the remaining $n-1$ annotators as follows: $\frac{1}{n}\sum_{i=0}^{n} | v_i - \frac{1}{n - 1}\sum_{j=0, j \neq i}^{n}v_j |$.

Across all values the LLM-Consensus MAE ($0.95\pm1.10$) is lower than the Human-Consensus MAE ($1.07\pm 1.05$). This result suggests that the LLM-generated labels agree with the consensus more than an individual human annotator agrees with the consensus. For all but three values (``Stimulation,'' ``Hedonism'', and ``Achievement''), the LLM performs as well as, if not better than, human annotators when it comes to predicting the consensus value label. For the three values where Human-Consensus MAE is less than the LLM-Consensus MAE, the difference between the two is at most $0.21$. See Appendix~\ref{sec:app_mae} for results disaggregated by value and additional evaluation metrics.

\subsection{Integrating multiple values together to rank content}
\label{section:ranking}
\begin{figure*}
    \centering
    \includegraphics[width=\linewidth]{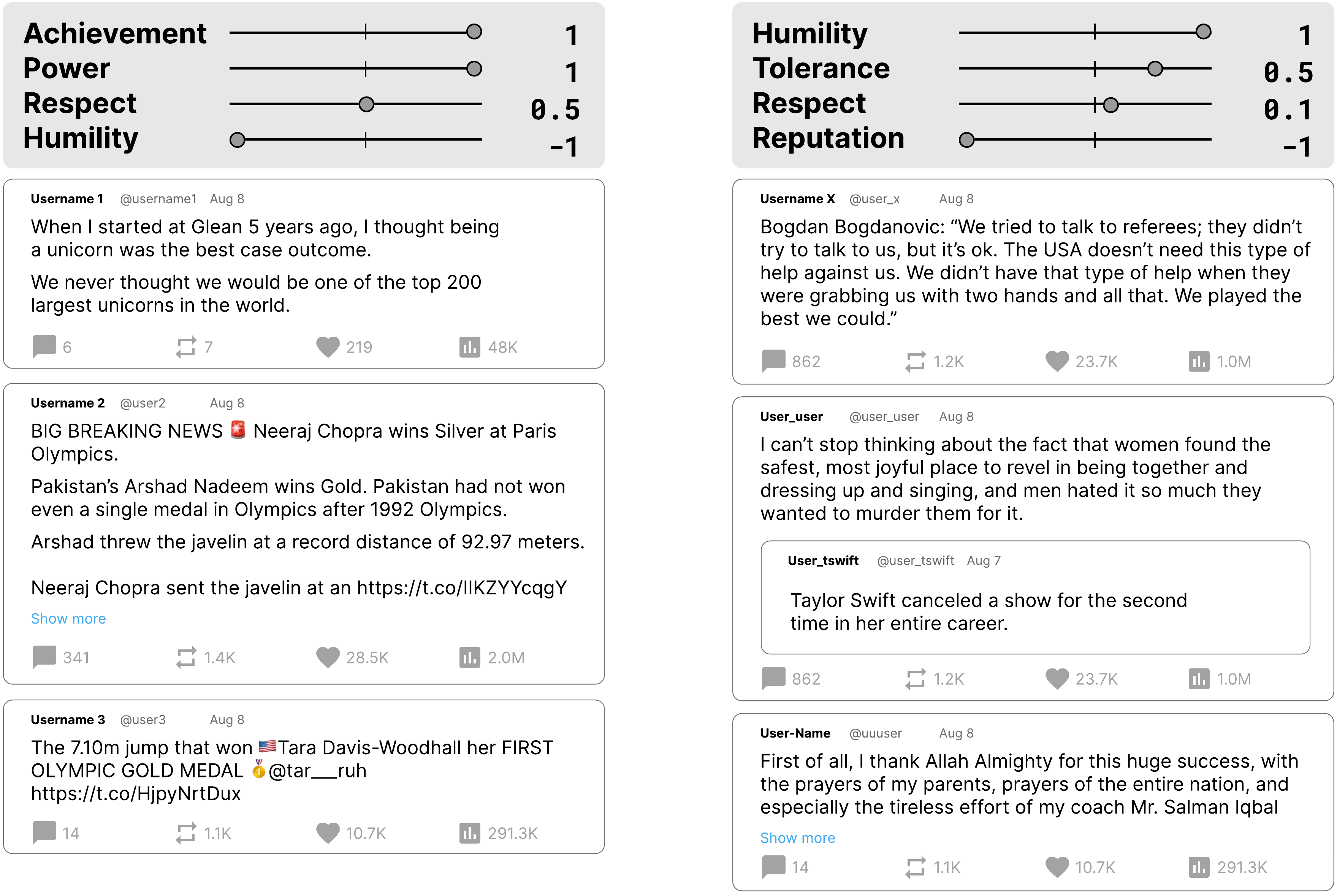}
    \caption{Our method integrates multiple values together when ranking content. We show the top posts from the same inventory of tweets but ranked by two different sets of value weights. The left shows a feed that prioritizes ``Achievement'' and ``Dominance'' (Power) while downranking ``Humility. The right feed prioritizes ``Humility'' while downranking ``Face'' (Reputation).}
    \label{fig:pudding}
    \Description[Side-by-side comparison of two simulated feeds showing the effect of different value weights.]{
    Side-by-side comparison of two simulated feeds showing the effect of different value weights. The left feed prioritizes “Achievement” and “Dominance” while downranking “Humility,” whereas the right feed prioritizes “Humility” while downranking “Face” (reputation), illustrating how ranking changes based on user-defined values
    }
\end{figure*}

Given content that is coded for presence and magnitude of values, we need a method for ranking content based on these values (Fig.~\ref{fig:pudding}). One approach would be to consider the individual's values as a strictly ranked list in which the values in higher positions should always be satisfied at the expense of those in lower positions. However, Schwartz argues that while individual values which are highly prioritized can be
predictive of behavior and attitudes, considering ``all'' values, especially tradeoffs between those that might be in conflict, provides a better representation~\cite{schwartz2013value}.

We take inspiration from Schwartz and devise a value-alignment score that depends on all activated values in a post. Our approach aims to integrate and navigate tradeoffs between values. We follow a common practice (e.g.,~\cite{narayanan2023}) of allowing the user to apply a linear weight with value between -1 and 1 to each value. Each weight dictates how much the final ranking should focus on that value. For example, if the ``Tradition'' value had a weight of $1$, and the ``Personal Security'' value had a weight of $0.25$, a post would need to have four times ($0.25 \times 4 = 1$) the classified score of ``Personal Security'' than a post with ``Tradition'' in order to rank equivalently. Many posts activate multiple values, so these weights become additive: for example, a post might have a weak expression of ``Tradition,'' a strong expression of ``Personal Security,'' and a strong expression of ``Societal Security.'' Then, the weights assigned to the values in the post add together.

Concretely, we re-rank feeds based on a set of weights, equivalent to how much the user does or does not want to see a given value, and the values assigned to each tweet from \texttt{GPT-4o}. We denote value labels as $\mathcal{V} = \{v_1, \dots, v_{n}\}$ for a set of $n$ tweets where $v_i$ is a one-dimensional row vector of length 19 taking values between $0$ and $6$. We also use a set of value weights, $w \in [-1, 1]^{19 \times 1}$, corresponding to whether a user wants to see more (i.e., $w_i > 0$) or less (i.e., $w_i < 0$) of a given value. For each tweet $i$, we obtain a score $w \cdot v_i$, with the dot product intuitively integrating the degree of alignment between the values the user wants to see and those present in the tweet's content. We then sort the scores in descending order.\footnote{In our current implementation, we do not consider the engagement scores or other features that determine the platform's default ranking. If desired, these engagement scores could be considered as one more weight in the calculation. However, in practice, our implementation only re-ranks content returned by the algorithmic ``For You'' feed, which means that all content has already passed through an algorithmic engagement filter at the platform level before our approach reranks it, so we found that, in practice, further accounting for engagement was unnecessary.} If there is a tie in scores, we retain the ordering from the original engagement feed. 

For example, suppose that one user, Jeanne, wants to rank their feed by Self-Transcendence values (e.g., ``Caring'', ``Dependability'', and ``Universal Concern''); another user, Jeff, wants to rank their feed by Self-Enhancement values that stand opposite of Self-Transcendence values (e.g., ``Achievement'', ``Dominance'', and ``Resources'') in Schwartz's circumplex. They each assign a weight vector $w$ with a weight of $1$ to the values they prioritize, $-1$ to the values on the opposite side of the circumplex, and leave all others at the default $0$ weight. Then, a tweet expressing strong support for ``Caring'' might get a score of $6$ from the classifier; ranking it highly for Jeanne as its score is $6 \times 1 = 6$, but at the bottom of the feed for Jeff, whose score is $6 \times -1 = -6$.

But few people are so simple in the values that they want to see represented in their feed. In practice, Jeanne may assign slightly different weights to each Self-Transcendence value, slightly reduce the weights for most --- but not all --- of Self-Enhancement values but increase the weight on ``Achievement'' because they wish to hear about their colleagues' accomplishments on social media. For each tweet in Jeanne's For You feed, our approach assigns numeric scores to every value expression, summing and weighing those scores based on Jeanne's weights.

Our approach is an example of how to navigate value elicitation and trade-offs, but is not the only way. Prior work has articulated other approaches such as reward learning from human feedback and metric elicitation, which offer algorithmic alternatives~\cite{christiano2017deep,hiranandani2019performance}. In contrast, we adopt a linear weighting scheme over values, which is a common choice in industry ranking pipelines~\cite{milli2023choosing,nytimes2021tiktok,twitter2023algorithm}. While linear weights cannot capture all potential interactions between values --- for example, preferences that depend on specific combinations of values --- they provide an interpretable starting point for enabling value-based ranking.

\subsection{Obtaining users' value preferences}
\label{sec:interfaces}

The final part of our pipeline is eliciting what values users may want to see (or not see) in their feeds: the weight vector $w$ on values. There is no single strategy for collecting these preferences. One method can be to learn value preferences from users' past behavior. Alternatively, users can directly input what values they want to see. Within this domain of end-user control, there are many options for how users can specify their values --- from having expert interfaces to simpler but less expressive controls --- that each come with their own trade-offs between afforded control and potential effort or cognitive burden. Our Study 1 below utilizes a heavyweight method of delivering a survey instrument to measure users' value priorities, then ranking feeds using the user's top values from their survey. However, it is possible that what users want to see in their feed may not align with their personal values. In Study 2, we offer users a set of sliders corresponding to each of Schwartz's values, offering more fine-grained control. Slider weights range from -1 to 1 in increments of 0.25. Since users may be unfamiliar with the values, we provide a tooltip next to each slider with the definition.

\section{Study 1: Re-ranking by a Single Value}
In Study 1, we aim to demonstrate that our method yields social media feeds that are recognizably ranked by a given set of basic human values. We start by evaluating whether our method succeeds when ranking by only a single basic value. We conducted a pre-registered experiment to evaluate whether participants are able to distinguish these value-aligned feeds from the current Twitter/X feed, which we refer to as an \textit{engagement feed},\footnote{Modern algorithmic feeds optimize for more than just engagement~\cite{narayanan2023}. However, Twitter's feed predominantly optimizes for engagement signals~\cite{twitterranker}.} at rates above random chance.\footnote{The study is pre-registered on OSF at this \color{ACMDarkBlue}\href{https://osf.io/7deug/?view_only=6d4f59207a5e4e73bbc307115d364943}{link}.} Through this study, we are able to validate our proposed ranking mechanism works and, furthermore, demonstrate that the value-aligned feeds can produce a meaningfully different experience for users. 

Our method hinges on the success of several components: basic values must be understandably conveyed to users; LLMs must accurately label value expression in tweets using these constructs; and the user's value weights must be accurately captured and reflected in the ranked feed. Furthermore, we feel that it is critical that we test success in the context of users' own social media feeds, as opposed to a pre-determined set of tweets. The distribution and nature of values present varies depending on the content in users' feeds, and an evaluation performed on crafted feeds may not have ecological validity; users may also not recognize the subtext or values in tweets outside their usual content exposure. Moreover, the evaluation should involve values that users find relevant or important to them, as users may struggle to recognize values they do not genuinely hold.


\subsection{Method}

\begin{figure*}
    \centering
    \includegraphics[width=0.8\linewidth]{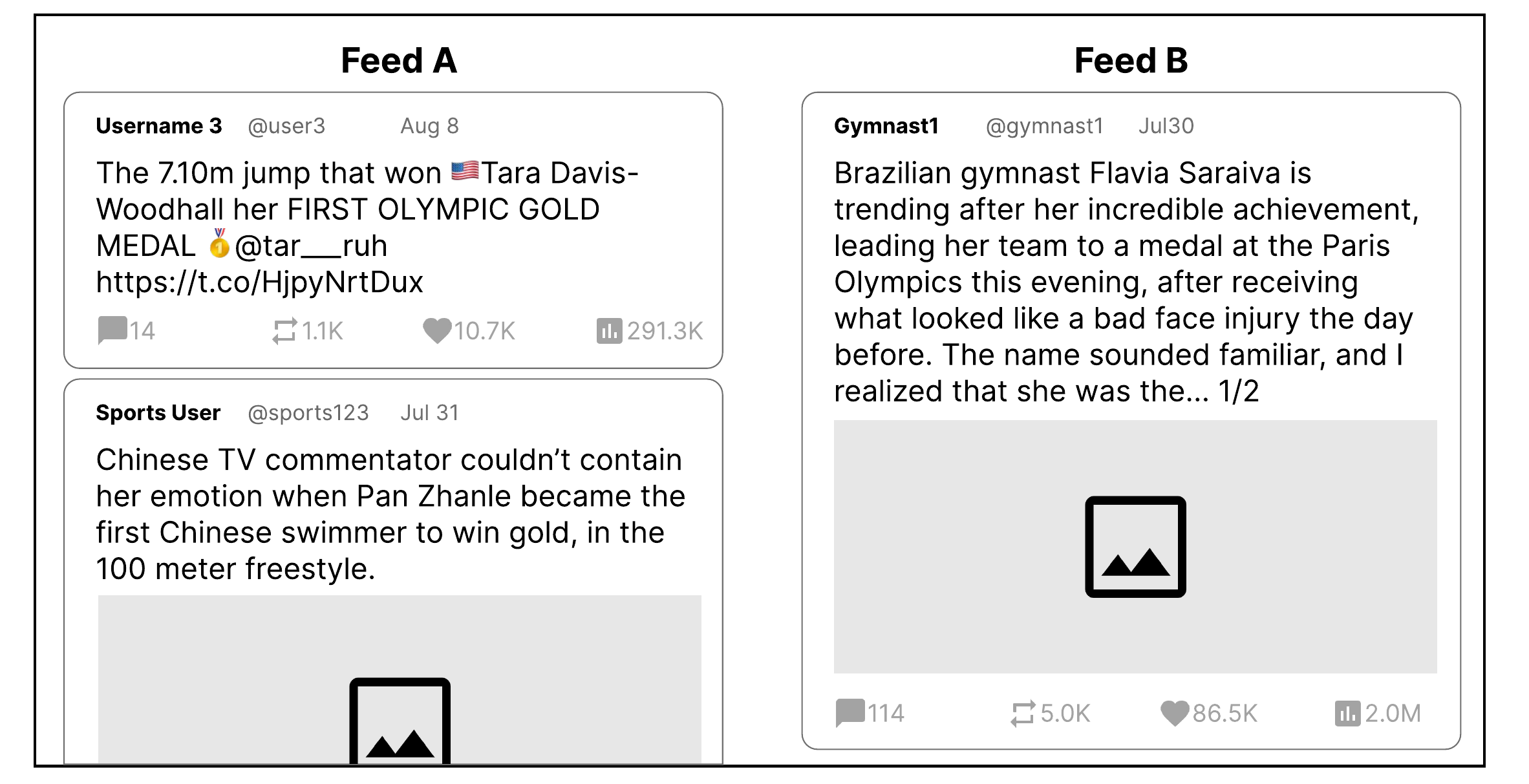}
    \caption{The experimental platform renders feeds emulating the style of Twitter. When comparing two feeds for recognizability, the engagement and value-ranked feed are shown side-by-side on the screen with generic labels (e.g., ``Feed A'') at the top of each feed.}
    \label{fig:feed_ex}
    \Description[Screenshot of the experimental platform interface.]{Screenshot of the experimental platform interface. Two unlabeled feeds are displayed side by side (“Feed A” and “Feed B”), visually replicating Twitter’s timeline style, used to test participants’ ability to distinguish between engagement-driven and value-aligned rankings}
\end{figure*}

We developed an experimental design in which participants ($N=141$) were repeatedly shown pairs of social media feeds and asked to select which of the two feeds was ranked by a specified value. To support both this experiment and our subsequent study (Sec.~\ref{sec:study2}), we created an experimental platform that renders the feeds in an environment designed to closely resemble a Twitter timeline (Fig.~\ref{fig:feed_ex}). 

\subsubsection{Study Design} 
Our study design has four steps. First, participants must download our browser extension so that we can collect an inventory of tweets from their own Twitter ``For You'' feed for us to use in the experimental task. After, participants are redirected to our experimental platform, where they complete an initial value survey. Third, in the main experimental task, participants compare two feeds and select the one they believe is ranked by a value they held strongly according to the value survey. There are four rounds of this task. Finally, participants complete a short demographic survey. We provide further details on each step below. 

In the first step, to collect tweets from the participants' Twitter feed, we ask them to download a browser extension that we have developed. The extension makes requests to the Twitter API on behalf of the user for fetching several consecutive batches of tweets. A ``batch'' is defined as the set of tweets returned from one call to the Twitter API, excluding those tweets that are ``promoted'' (advertisements) and recommendations about who to follow or to subscribe to. We collect our batches from the user's ``For You'' timeline (the engagement feed). By using tweets from the engagement feed, we can control for user interest and satisfaction with the content. It also provides a more realistic window into how social media companies may adopt value-based ranking mechanisms on top of existing algorithms. These tweets are the inventory of content being ranked on our platform. Although the content on users' ``For You'' timelines varies, leading to differences in the posts that participants see, we chose to use posts from participants' own feeds rather than a static set across users, as this setup more realistically reflects real-world feed ranking deployment.

Participants are then redirected to our external experimental platform website, where they complete the Portrait Value Questionnaire (PVQ)~\cite{schwartz2012overview}, a 57-item survey used to measure an individual's prioritization of the 19 Schwartz values. We use the results from this survey to determine the weights for the value-aligned feed. In each trial for Study 1, we test how well our approach can rank by a single value at a time. To do so, we set the weight of one value equal to 1, and the weight of all other values to 0. Specifically, for the first trial, we select only the participant's highest-ranked value from the PVQ, subsetting to only the values for which there is at least one tweet in the inventory expressing this value. Next, in the second trial, we look at the second highest-ranked value that also appears in our inventory, and so on. Each participant thus sees a set of values that reflect their top-ranked values as rated in the PVQ. There are four such trials, each focused on a different highly-rated value for that participant.

In the main task of the study, users are shown two side-by-side feeds and are asked, ``Which feed contains more content reflecting \{VALUE\}?'', where the value was one of their top four PVQ values as described above. We also provide a definition of the value. Concretely, participants see two side-by-side feeds on one screen --- one of which is the value-aligned feed and the other has Twitter's engagement ranking. The right or left placement for the two feeds is determined randomly, and the feeds are not labeled as being engagement- or PVQ-ranked. 

For each trial, we draw a fresh sample of up to nine sequential batches (roughly 270 tweets) from the participant's For You feed. We order the tweets according to the ranking for each condition (value-based or engagement), then we select the top twenty from that ranking to create the feed. We only show twenty posts to reduce participants' cognitive burden. Therefore, although the inventory of tweets used to construct the two feeds is the same, the content displayed may differ.

\subsubsection{Participants}
We recruited participants in March 2025 via the Prolific platform. To qualify for our studies, participants must reside in the United States, be over 18 years old, have a Twitter account, and pass quality checks on Prolific (i.e., over 250 prior submissions and $\geq95$\% acceptance rate). Since values may differ depending on individuals' political orientation ~\cite{schwartz2012political,haidt2004intuitive, graham2009liberals}, we sampled participants from across the political spectrum to ensure we are not systematically excluding any values. We launched separate tasks on Prolific to recruit participants who self-identified as Democrats and for those who self-identified as either Republican or Independent.\footnote{We combined recruitment for Republicans and Independents to ensure adequate sample size.} Participants received \$4.25 for completing the task, and those that correctly identified all four value re-ranked feeds received a bonus of $\$0.50$. We determined the payment from pilot studies and our intention to provide a rate of \$15 per hour. In total, we collected responses from 194 participants and removed those that failed our attention checks ($N=17$) or provided incomplete responses ($N=36$). For our analyses, we were left with 564 responses from 141 participants. 

\subsubsection{Measures} Our main measure of interest in Study 1 is \textit{recognizability}. In other words, we are interested in whether users are able to distinguish of the two presented feeds is ranked by the specified value. We measure the rate of recognizability as the percentage of trials for which the value-aligned feed is correctly selected out of the total number of trials.


\subsection{Results}

\subsubsection{Participants are able to recognize value-aligned feeds} 
We find that participants can correctly distinguish feeds ranked using values at rates well-above random chance. In $76.1$\% of trials, participants identified the value-aligned feed; in the remaining $23.9\%$ of trials, they mis-identified the engagement-aligned feed as the value-aligned feed. Furthermore, $36.8\%$ ($N=52$) of participants were able to correctly identify all value-aligned feeds shown to them. To formalize this analysis, we conduct a chi-square goodness of fit test. We compare our observed distribution to an expected split of $50\%$, which would suggest there are no distinguishable features between the feeds (i.e., participants are selecting at random). The chi-square test results ($\chi^2=153.26$, $p < 0.001$) indicate there is a significant difference in our observed proportions from the random-chance scenario. 

To account for the fact that our study has repeated measures, we also report the results of fitting a mixed-effects logistic regression to our data with a random effect for \texttt{Participant}. We use an intercept-only model to predict our dependent variable, \texttt{Correctly Recognized}, which is a binary variable equal to 1 if the participant selected the value-aligned feed and 0 otherwise. The model's intercept of $1.231$ represents the log-odds of selecting the value-aligned feed or equivalently a probability of $0.774$. 

\subsubsection{Most --- but not all --- values produce recognizable feeds} 
\begin{table*}[]
    \centering
    \begin{tabular}{llrrr}
    \toprule
    \textbf{Schwartz Quadrant} & \textbf{Value} & \textbf{N} & \textbf{Recognizability (\%)} & \textbf{95\% CI} \\
    \midrule
\textcolor{se}{\textbf{SE}} & Dominance & 4 & 100.0 & [100.0, 100.0] \\
\textcolor{c}{\textbf{C}} & Rule conformity & 16 & 93.8 & [81.2, 100.0] \\
\textcolor{st}{\textbf{ST}} & Preservation of nature & 10 & 90.0 & [60.0, 100.0] \\
\textcolor{st}{\textbf{ST}} & Caring & 39 & 89.7 & [79.5, 97.4] \\
\textcolor{st}{\textbf{ST}} & Universal concern & 55 & 89.1 & [81.8, 96.4] \\
\textcolor{st}{\textbf{ST} / \textcolor{c}{\textbf{C}}} & Humility & 15 & 86.7 & [66.7, 100.0] \\
\textcolor{se}{\textbf{SE}} & Achievement & 30 & 86.7 & [73.3, 96.7] \\
\textcolor{c}{\textbf{C}} & Tradition & 17 & 82.4 & [64.6, 100.0] \\
\textcolor{st}{\textbf{ST}} & Tolerance & 11 & 81.8 & [63.4, 100.0] \\
\textcolor{se}{\textbf{SE}} / \textcolor{c}{\textbf{C}} & Face & 20 & 80.0 & [60.0, 95.0] \\
\textcolor{st}{\textbf{ST}} & Dependability & 29 & 79.3 & [62.1, 93.1] \\
\textcolor{c}{\textbf{C}} & Societal security & 45 & 75.6 & [62.2, 86.7] \\
\textcolor{oc}{\textbf{OC}} & Self-directed thoughts & 84 & 72.6 & [63.1, 81.0] \\
\textcolor{c}{\textbf{C}} & Personal security & 36 & 69.4 & [55.6, 83.3] \\
\textcolor{se}{\textbf{SE}} / \textcolor{oc}{\textbf{OC}} & Hedonism & 39 & 69.2 & [53.8, 82.1] \\
\textcolor{oc}{\textbf{OC}} & Stimulation & 18 & 66.7 & [44.4, 88.9] \\
\textcolor{se}{\textbf{SE}} & Resources & 8 & 62.5 & [25.0, 100.0] \\
\textcolor{oc}{\textbf{OC}} & Self-directed actions & 71 & 62.0 & [50.7, 73.2] \\
\textcolor{c}{\textbf{C}} & Interpersonal conformity & 17 & 47.1 & [23.5, 70.6] \\
    \bottomrule
    \end{tabular}
    \caption{On average most values can produce recognizable feeds. We disaggregate recognizability by each of Schwartz's 19 values and include the broader group that each value belongs to using the following keys (ST = Self-Transcendence, OC = Openness to Change, SE = Self-Enhancement, and C = Conservation). We report the number of participants ($N$) for which the feed was ranked using the given value as well as $95\%$ confidence intervals bootstrapped over 1,000 iterations.}
    \label{tab:disagg}
\end{table*}

We also conducted a disaggregated analysis to examine recognizability across the 19 values. We find that for all but one of the values (``Interpersonal Conformity''), mean recognizability is above $50\%$ (Table~\ref{tab:disagg}). We also compute $95\%$ confidence intervals by bootstrapping over 1,000 iterations and find that all but three (``Interpersonal Conformity'', ``Resources'', and ``Stimulation'') have recognizability rates significantly higher than random chance. We observe a wide range of recognizability, spanning from $100.0$\% for ``Preservation of Nature'' to $47.1\%$ for ``Interpersonal Conformity''. Values grouped under the category Openness to Change (e.g., ``Self-directed actions'' and ``Stimulation'') had relatively lower recognizability, with a mean of $67.6\%$, compared to the Self-Transcendence group (e.g., ``Caring'' and ``Humility''), which has a mean recognizability of $86.1\%$. One explanation for this variance may be the interpretability of different values. Even though we translated Schwartz's constructs into more understandable terms, certain constructs are easier to convey than others. For example, understanding what ``Caring'' means, which scores higher on recognizability ($89.7\%$), may be easier for participants who are unfamiliar with Schwartz's Basic Values compared to a construct such as ``Self-directed actions'' ($62.0\%$). Another factor may be because the underlying engagement ranking already filters content, implicitly selecting for certain values, such as those related to Openness to Change. As a result, feeds ranked by these values may have produced less pronounced differences.

\subsubsection{Summary} Participants are able to distinguish feeds that are ranked using a single human value, providing supporting evidence that our method can execute recognizable rerankings of a social media feed. While the rate of recognizability does differ depending on which of the 19 values are used to rank the feed, the majority of values are recognizable. This first study constrains our ranking algorithm to consider only one value at a time. Within this constraint, we confirm that our approach successfully embeds human values into a feed-ranking algorithm. For our participants, these resulting feeds are meaningfully different from the status-quo engagement ordering. Study 2 will loosen several of these constraints, allowing participants to author their own desired feeds and control multiple values simultaneously.

\section{Study 2: Re-ranking by Multiple, User-Controlled Values}
\label{sec:study2}
\begin{figure*}
    \centering
    \includegraphics[width=\linewidth]{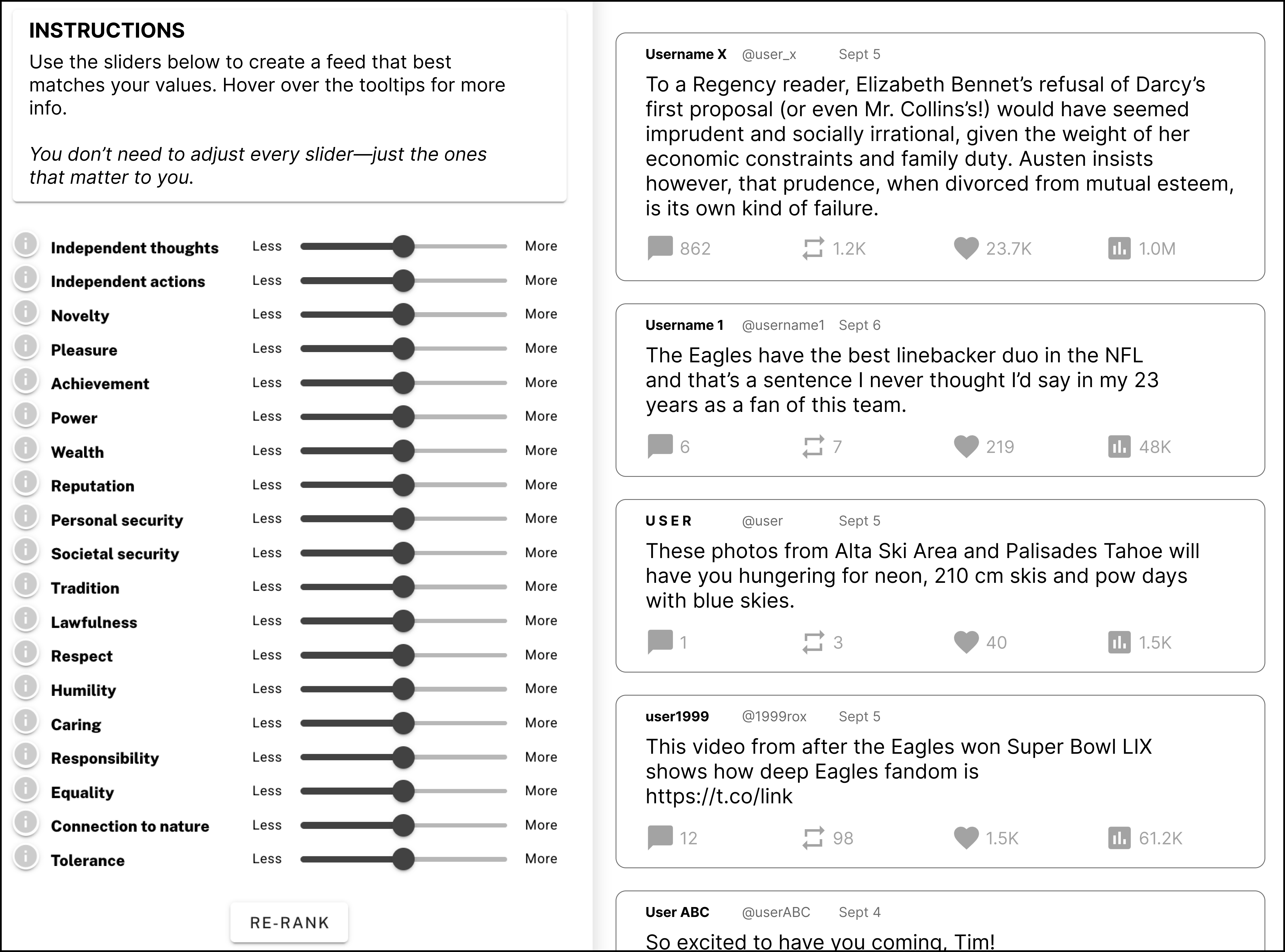}
    \caption{On our experimental platform, participants are able to re-rank their feed by adjusting value sliders. Participants see a single feed on the right-side of their screen presented with the sliders that they can use to re-rank the content on the left-side. While participants are shown all available sliders, the number that they can adjust (i.e., 1, 2, 3, 4, 5, or all) depends on their assigned condition.}
    \label{fig:interface}
    \Description[Screenshot of the interface for Study 2]{
    Screenshot of the interface for Study 2, where participants could adjust slider weights for up to 19 values. The left side shows the feed being re-ranked in real time, while the right side presents the 19 sliders.
    }
\end{figure*}

In Study 2, we evaluate whether our ranking method works when users are given control over specifying the values they want to see in their feeds, using a pre-registered experiment.\footnote{The study is pre-registered on OSF at \color{ACMDarkBlue}\href{https://osf.io/tm4uj/?view_only=cfd440aebc7d4da8bbbbd52716cfd048}{link}.} Whereas Study 1 validated that our method performs well when ranking by a single value, in practice, users often hold multiple values they simultaneously want to prioritize or avoid in their feeds. However, when multiple values are involved, satisfying all preferences becomes challenging due to inherent trade-offs between competing priorities. This added complexity makes it a nontrivial task for users to recognize whether a feed has been successfully re-ranked according to their intentions.

Furthermore, multi-value ranking introduces an important design question: how many values can users reason over and recognize at once? Increasing the number of adjustable parameters provides users with more granular control but may also make it cognitively harder to evaluate multi-objective trade-offs. The answer to this question informs interface design: how many value dimensions should a system expose to users at once? This study, therefore, aims to investigate two core questions:

\begin{enumerate}
\item Are users able to accurately recognize feeds re-ranked by values that they choose themselves?
\item At what limit, if any, does recognizability suffer as the number of simultaneous values a user adjusts increases?
\end{enumerate}

\subsection{Method}
We conducted a between-subjects experiment (N=250) eliciting participants' value weights using a set of sliders (see Fig.~\ref{fig:interface}). Participants were assigned into one of six conditions. While participants were shown sliders corresponding to all nineteen values, their assigned conditioned determined the maximum number of sliders they could adjust. The number of value sliders participants could adjust ranged from 1 to 5 or all 19 value sliders (i.e., ``Full'' condition). Our pilot tests, described in more detail in Appendix~\ref{sec:pilot}, included conditions for adjusting 1 to 6 values as well as the full 19. The results suggested that recognizability began to plateau beyond just a few values and informed our pre-registered final study, where we included 1 to 5 sliders and all 19.

\subsubsection{Study Design}
The user flow in Study 2 closely resembles that of Study 1. After a user installs our browser extension, the extension fetches content from their ``For You'' feed on Twitter and redirects them to our experimental platform. Once on the platform, the user first completes the PVQ, then proceeds to the experimental task. Following the task, they complete a demographic survey. 

In the experimental task, the user is exposed to a ``training'' and ``testing'' phases. During the training phase, the participant uses the controls to create a feed that best matches their values. Depending on the condition, participants are able to change only a select number of value sliders (or all if they are in the Full condition). They must change at least one value to proceed. After the training phase, the user answers a survey on the perceived complexity of the interface as well as providing a free-text response on what they liked and did not like about the interface. In the testing phase, similar to Study 1, users are shown a side-by-side comparison of their value-aligned feed and engagement feed (i.e., the feed as ranked by Twitter/X's algorithm), without labels and in a randomized position. As a robustness check, we replicated the conditions in which participants could adjust either a single slider or all 19, but instead asked them to choose between the value-aligned feed and a randomly reshuffled feed. We found qualitatively similar results (see Appendix~\ref{sec:recog_robustness} for full details). They are asked to identify which feed has been ranked using the weights from their input across four trials. In each trial, the tweets shown to the participant come from a different set of batches. For each trial, the pool of possible tweets being ranked consists of up to seven batches (roughly 210 tweets), and each feed shows only the top 20 ranked tweets by its own ranking function. 

\subsubsection{Participants} We recruited participants from the Prolific platform in July 2025. Participants must reside in the United States, be over 18 years old, have a Twitter account, and pass quality checks on Prolific (i.e., over 250 prior submissions and $\geq95$\% acceptance rate). Again, we sampled participants from across the political spectrum following the protocol described in Study 1. We compensated each participant with \$5.00, at an hourly rate of \$15. Participants who correctly identified all four value re-ranked feeds were given a bonus of $0.50$. We preregistered a sample size of 240 participants or equivalent to 40 participants in each condition. We recruited $272$ participants, filtering out $22$ participants who failed our attention checks. After excluding these participants, we were left with a sample size of $250$ participants whose responses were included in our analysis. 

\subsubsection{Measures} We focus on two main measures in Study 2: recognizability and task load. Similar to Study 1, \textit{recognizability} is measured as the percentage of trials in which participants correctly select the value-aligned feed in an unlabeled side-by-side comparison (Fig.~\ref{fig:feed_ex}). In Study 2, we add an additional dependent variable, \textit{task load}. To measure task load, we adapt questions from the NASA Task Load Index (TLX)~\cite{hart1988development}, participants rated how hard (``I had to work hard to create the resulting feed.'') and mentally demanding (``Using the controls was mentally demanding.'') each interface was to use on a 7-point Likert scale (1=``Strongly disagree'', 7=``Strongly agree''). We average over the responses as our measure of task load (Cronbach's $\alpha = 0.66$). The survey measures are gathered after the training round but before the testing rounds.

\subsection{Results}
Again, we find that participants are able to consistently recognize feeds re-ranked based on one value at a rate of $72.6\%$. This task is more difficult than that in Study 1, which is equivalent to a setting where a participant upranks a single value with a weight of 1. In the experimental setup in Study 2, participants still choose only one value, but they can choose to uprank or downrank and can adjust the weight of the value to be range between -1 and 1. Having established that feeds can be effectively re-ranked in this more naturalistic setting using a single value, we next push the boundaries of our design by investigating re-ranking when multiple values interact simultaneously.

\begin{figure}
    \centering
    \includegraphics[width=\linewidth]{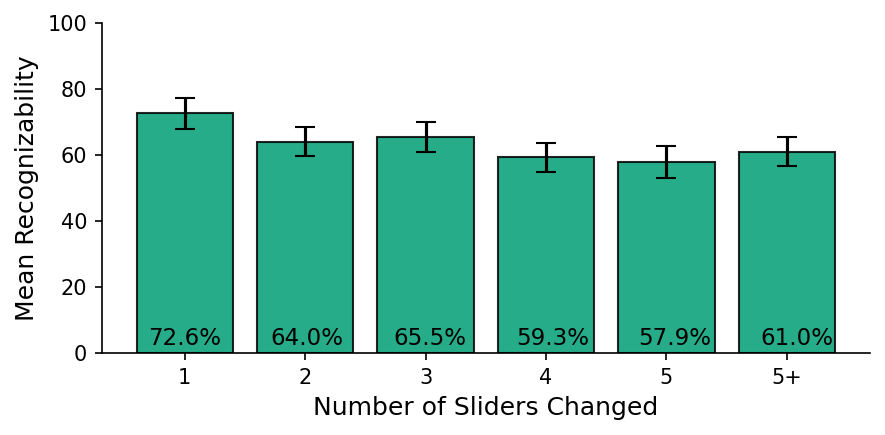}
    \caption{Recognizability drops as the number of sliders changed increases but remains above random chance even when participants are able to change all 19 sliders. We group together conditions with more than 5 sliders changed into a single ``5+'' category due to the small number of participants in higher-change conditions. All error bars represent 95\% confidence intervals.}
    \label{fig:study2_changed}
    \Description[Bar chart showing recognizability rates by number of sliders adjusted in Study 2.]{
    Bar chart showing recognizability rates by number of sliders adjusted in Study 2. Recognizability declines as more values are simultaneously activated, from 72.6\% for single-value ranking to 64.0\% for 2 sliders, 65.5\% for 3 sliders, 59.3\% for 4 sliders, 57.9\% for 5 sliders, and 61.0\% for 5+ sliders.
    }
\end{figure}

\subsubsection{Value aligned feeds become harder to recognize as more values are activated} 
While overall recognizability remains well above random chance at $63.4\%$, we find that recognizability decreases for feeds re-ranked using multiple values compared to those re-ranked using only one. We formalize this analysis using a mixed-effect logistic regression model with \texttt{Participant} as a fixed random effect and {Correctly Recognized} as binary dependent variable (1 if the participant selected the value-aligned feed during the testing round and 0 otherwise). Our key predictor is \texttt{Sliders Changed}, which captures the number of sliders that participants adjusted.\footnote{\texttt{Sliders Changed} differs from the assigned condition, since participants could adjust fewer sliders than the maximum available. Results are qualitatively similar when we use the assigned condition as predictor variable. See Appendix for results.} 

Overall, we observe a weak negative association between the number of sliders changed and recognizability ($\beta = -0.028$), but this effect is not statistically significant ($p = 0.14$). As shown in Fig.~\ref{fig:study2_changed}, a closer inspection reveals a noticeable drop in recognizability when comparing participants who changed exactly one value ($72.6\% \pm 30.1$) to those who changed more than one ($61.5\% \pm 29.1$). To formalize this observation, we conduct an exploratory analysis, replacing \texttt{Sliders Changed} in our linear mixed-effect model with a binarized predictor \texttt{>1 Slider Changed}. \texttt{>1 Slider Changed} is equal to 1 when the number of sliders changed is greater than 1 and 0 otherwise. We observe a significant negative correlation between \texttt{>1 Slider Changed} and recognizability ($\beta=-0.60$, $p = 0.02$). In short, recognizing value aligned feeds becomes more difficult when multiple values are activated, likely because participants are forced to reason over more concepts each exerting different degrees of influence on the feed, increasing their cognitive load~\cite{popowski2026people}. However, beyond the initial drop-off from re-ranking using a single to multiple values, recognizability does not decline significantly as the number of value dimensions increases.

\subsubsection{Task load does not increase across conditions}
We also examined whether participants perceived the interface as being more complex when they are able to adjust more sliders. Participants in the single-slider condition reported the lowest perceived task load (M = $1.42$, SD = $1.27$) compared to conditions where they could adjust multiple sliders. However, independent $t$-tests comparing all six conditions, with Bonferroni correction applied to control for multiple comparisons, revealed no statistically significant differences in reported task load. Qualitative feedback provides additional insight into this pattern. Many participants described the cognitive effort involved in prioritizing which values to adjust, noting that it was challenging to decide which sliders mattered most (e.g., ``\emph{it was difficult to have to choose only two max that I could adjust}'') or expressed frustration about needing to limit their changes (``\emph{there were a lot of sliders that I felt were important to me, so it was tough only choosing five}''). Conversely, some participants in the Full Sliders condition reported experiencing choice overload, remarking that there ``\emph{were too many options to choose from}'', leading them to feel overwhelmed by choice fatigue.

\section{Understanding Users' Value Selections}
Finally, we investigate how and when users want to use value-based controls to curate their feeds. While our earlier studies examined whether value-based re-ranking can produce recognizable feeds, here we focus on how participants make sense of these controls and in what contexts they find them useful. We conduct a quantitative analysis of the value selections and resulting feeds for the 47 participants from Study 2 who were assigned to the Full Sliders condition.

\subsection{What values do users select to re-rank their feeds?}
A confluence of factors --- from the values that are actually present in the available content to users' mood --- can influence what values users select. Because these selections are highly individualized, our analysis focuses on uncovering the mechanisms driving these choices rather than merely characterizing aggregate selection patterns.

\noindentparagraph{Configured value weights reflect personal values.} One hypothesis is that participants adjust sliders to reflect their personal value priorities. To test this, we compute the Pearson correlation ($r$) between each participant’s personal value rankings, measured via the Portrait Value Questionnaire (PVQ), and their selected slider weights. For each of the nineteen values, the PVQ returns a continuous score ranging from $-5$ to $5$, which represents the relative importance of the value to the individual. The results support our hypothesis that participants' slider selections match their personal values: $93.6\%$ of participants’ selected sliders are positively correlated with their personal values, with a mean correlation of $r = 0.39 \pm 0.25$. While only $38.3\%$ of participants show individually significant correlations ($p < 0.05$), a one-sample $t$-test on Fisher’s $z$-transformed coefficients confirms that, overall, the positive alignment is highly significant ($p < 0.001$).\footnote{Since each participant's correlation is based on at most 19 values, individual tests are underpowered. Thus, many positive correlations do not reach the significance threshold despite being moderate in size.} 

\noindentparagraph{Value weights mirror Schwartz's trade-off structure.} We select Schwartz's value system as it presents a design space of values containing trade-offs, and enumerates values that are complementary versus those in tension in a structured fashion (i.e., based on distance in the circumplex). If users' value selections reflect these theoretical relationships, we would expect that users are unlikely to simultaneously uprank values that are in tension or downrank values that are complementary. To examine how these trade-offs manifest in users' configured value weights, we compute the Pearson correlation ($r$) between the distance separating two values in Schwartz's circumplex and the product of the weights the user assigned to them. We define distance $d$ as $1 - \frac{2\theta}{\theta_{\max}}$ where $\theta$ is the angular distance between two values in the circumplex (Fig.~\ref{fig:circ}) and $\theta_{\max}$ is the maximum possible angular distance. This metric ranges from -1 for values at the opposite ends of the circumplex to 1. After applying a Fisher $z$–transformation, we observe a significant negative relationship between our distance metric and value weights with a mean correlation coefficient of $-0.15$ ($p < 0.001$). This result indicates that users tend to assign higher weights to complementary values (e.g., ``Caring'' and ``Universal concern'') and lower weights to values in tension (e.g., ``Tradition'' and ``Self-directed thought''), consistent with engaging in meaningful value trade-off reasoning.

\subsection{How do value-aligned feeds compare to engagement feeds?} Next, we consider how participants' selections actually impact the feeds that they see. Are we slightly adjusting the engagement-based ordering, or do value-based feeds contain completely different content? We start by measuring the strength of values present in the existing engagement-aligned feed before analyzing how value-aligned feeds differ.

\noindentparagraph{Engagement feeds prioritize individualistic values.}
\label{subsec:dcg}
\begin{figure*}[h]
    \centering
    \includegraphics[width=0.9\linewidth]{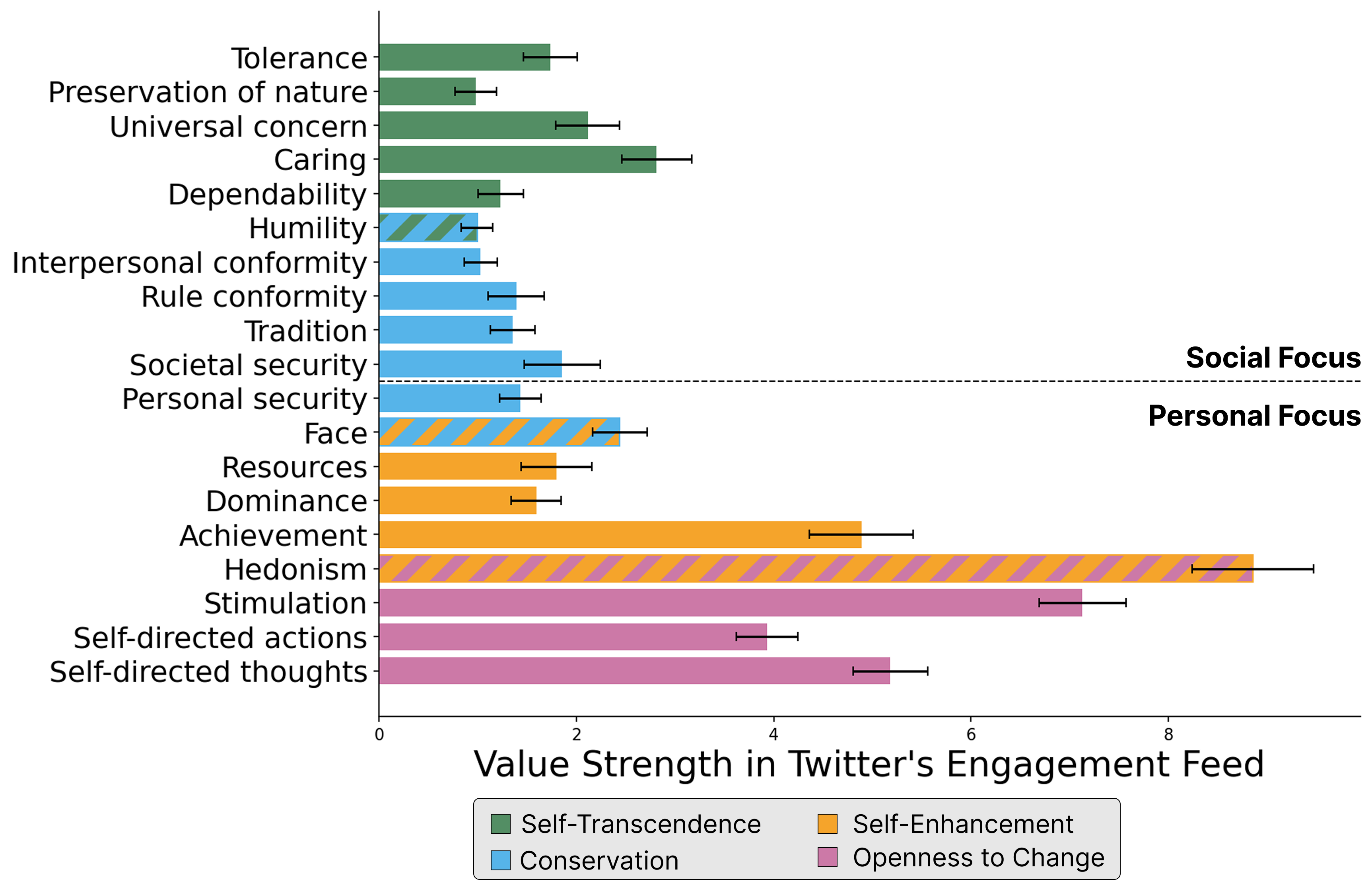}
    \caption{Engagement feeds prioritize individual or personal values over societal ones. We visualize the average strength of each value across participants' engagement feeds. Value strength is calculated by taking the sum of the label for the value discounted by the tweet's position in the feed. The colored bars indicate the broader group each value belongs to --- hatched bars indicate that the value belongs to two groups. The dotted horizontal line separates values that Schwartz defined as having a social focus versus a personal focus~\cite{schwartz2012refining}. Error bars represent $95\%$ confidence intervals.}
    \label{fig:strength}
    \Description[Bar chart showing the average strength of each value in participants’ engagement-aligned feeds.]{
    Bar chart showing the average strength of each value in participants’ engagement-aligned feeds. Values with a personal focus, such as "Stimulation", "Achievement", "Hedonism", appear more frequently than social-focus values, like "Caring" or "Humility."
    }
\end{figure*}

We find that values corresponding to the individual and outcomes for the self are featured prominently (Fig.~\ref{fig:strength}).  We calculate the value strength across all of participants' engagement feeds. Value strength is the sum of the value labels returned by the LLM, discounted by the position of the tweet in the feed. In line with existing literature (e.g., DCG)~\cite{jarvelin2002cumulated}, we add this discounting factor to account for the fact that values shown in tweets at the top are more salient and should thus be weighted higher. Concretely, given a set of $n$ tweets denoted by $\mathcal{T} = \{t_0, t_1, \dots, t_n\}$, and a model $\mathcal{X}$, we calculate value strength as $\sum_{i=0}^{n}\mathcal{X}(t_i) / \log_2{(i+2)}$. Values with a personal focus appear more strongly ($4.14\pm3.70$) in feeds compared to those with a social focus ($1.55\pm1.97$). Across the 19 values, the most strongly present values are ``Hedonism'' ($8.86\pm4.31$) and ``Stimulation'' ($7.13\pm3.08$) . These results corroborate prior work~\cite{bernstein2023embedding}, which claims that optimizing for engagement will prioritize individualistic values. 

\noindentparagraph{Value-aligned feeds are uncorrelated with engagement-based rankings} How do the feeds that participants configure compare to the existing engagement feeds? We calculate the rank correlation (Kendall's $\tau$) between the two. The mean Kendall's $\tau$ for all the interfaces is $0.06\pm0.14$, with values ranging from $-0.57$ to $0.67$. The ordering of the value-ranked feed is uncorrelated with that of the engagement feed, indicating that current ranking algorithms do not reflect the values that users want to see in their feeds.

\noindentparagraph{Self-transcendence and openness to change values are amplified.} 
\begin{figure*}[h!]
    \centering
    \includegraphics[width=0.9\linewidth]{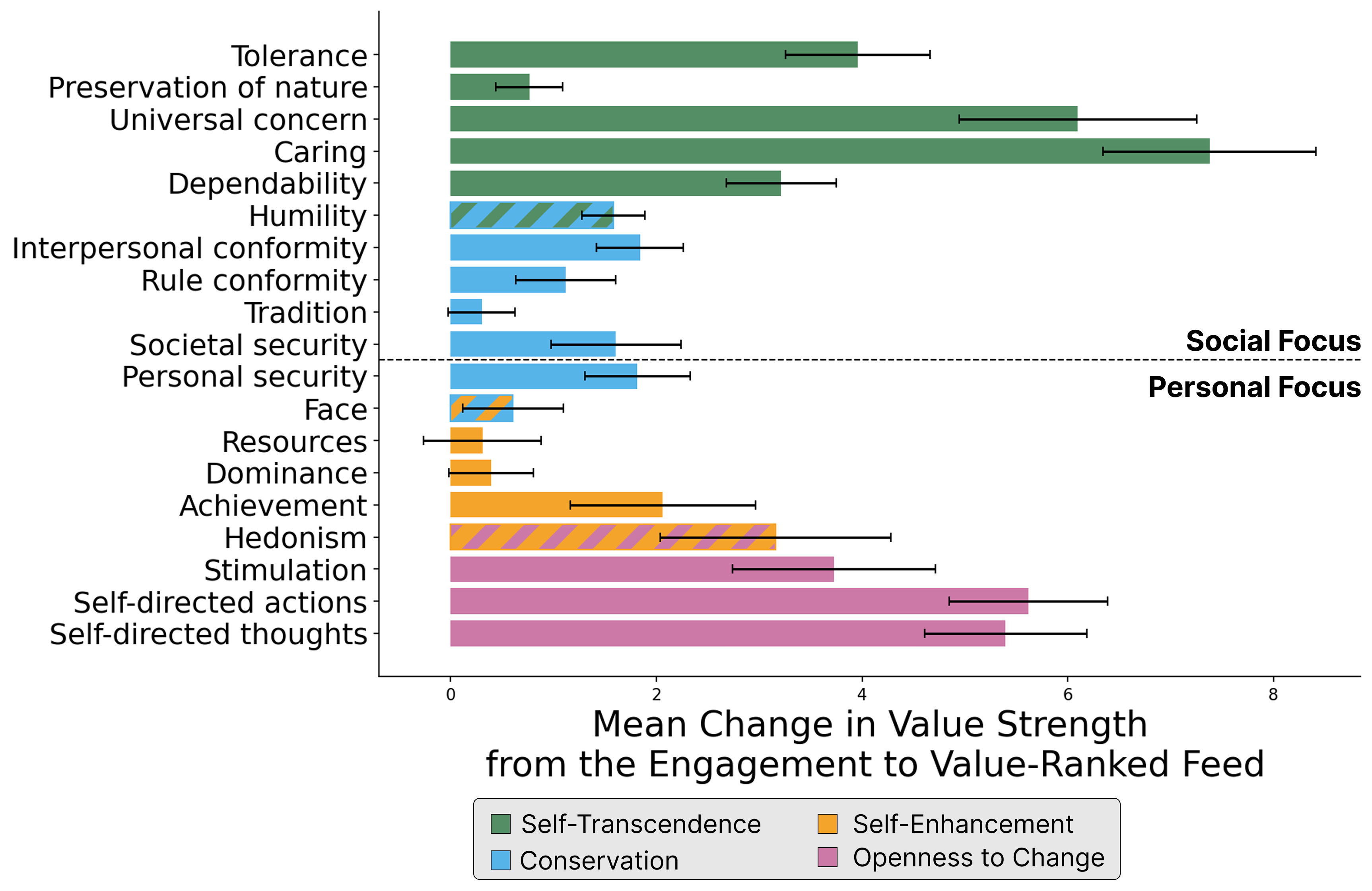}
    \caption{Content containing values related to self-transcendence and openness to change are amplified in value-ranked feeds. The colored bars indicate the broader group each value belongs to --- hatched bars indicate that the value belongs to two groups. The dotted horizontal line separates values that Schwartz defined as having a social focus versus a personal focus~\cite{schwartz2012refining}. Error bars represent $95\%$ confidence intervals.}
    \label{fig:perc_change}
    \Description[Bar chart showing the mean change in value strength from participants’ engagement-aligned feeds to the value-aligned feed.]{
    Bar chart showing the mean change in value strength from participants’ engagement-aligned feeds to the value-aligned feed. There is a greater increase for values related to Self-Transcendence (e.g., "Caring", "Universal Concern") and Openness to Change ("Self-directed actions", "Self-directed thoughts").
    }
\end{figure*}

Finally, we analyze what values are being promoted versus deprioritized in the re-ranked feeds. Using the method from Sec.~\ref{subsec:dcg},  we compute the change in value strength from the participant's engagement feed to the value-ranked feed for each testing round (Fig.~\ref{fig:perc_change}). Here, the maximum possible change in value strength is $42.4$, which is equivalent to going from an engagement feed where the value is not expressed to a re-ranked feed where each post strongly expresses the value. Overall, value strength increases across all dimensions as more value-laden content is promoted during re-ranking. There is a larger boost for values expressing self-transcendence ($4.47\pm1.053$) as opposed to self-enhancement ($1.27\pm1.13$) and openness to change ($3.83\pm2.33$) over conservation ($1.31\pm1.13$). Value-aligned feeds contained more content concerning the welfare or interests of others rather than posts emphasizing the pursuit of individuals' interests.

\section{Discussion}
Much of the content that appears on our feeds is loaded with values: some call for unity, others promote uniqueness; some embrace change, others uphold tradition. The influence that these values have on us is not confined to the online realm. They shape our perceptions of the world and consequently how we interact with it. And so it seems inadequate that we cannot control these values, that their presence or absence in our feeds should be a second degree effect of optimizing for surface-level engagement signals.

Our work envisions a social media ecosystem where basic human values, with their inherent trade-offs, are directly incorporated into feed ranking algorithms. In this section, we highlight both the promise and the complexity of building value-aligned social media feeds, drawing on the design decisions we worked through. Designing real systems that support coherent value alignment requires design across the entire pipeline, from how to design interfaces that enable users to express their values (Sec.~\ref{subsec:interface}), to how values could be incorporated into ranking infrastructure (Sec.~\ref{subsec:ranking}), to how the system interprets and classifies values in content (Sec.~\ref{sec:democratized}). Below we outline several design considerations for building such systems in practice.

\subsection{Eliciting Value Priorities}
\label{subsec:interface}
The interface we designed to elicit users' value priorities used 19 sliders, one for each value. Yet this approach represents just one point in a larger design space. Exploring alternatives requires grappling with trade-offs between interface complexity, users' perceived control over their feed, and the recognizability of the resulting feeds. One promising direction is eliciting values via natural language, which provides users with greater expressivity. For example, users might want for values to be activated conditionally or in combination with each other, such as wanting to see content that expresses only societal security and caring in political posts. This level of expressivity is not possible with our current interface but could be elicited via natural language. An open question for natural language interfaces is how to best guide users' reflection, such as through free-form text entry or a multi-turn interview setting~\cite{li2025eliciting}.

A broader challenge in interactive value elicitation is that users must anticipate how their selections will translate into ranking behavior. This task becomes even more challenging when users must reason over how multiple values interact with each other, as the feed may be influenced in different, and sometimes, conflicting directions. Our value elicitation interface in fact masked some of these conflicts by allowing users to adjust each value independently. Nonetheless, participants could still distinguish which feeds were aligned with their intentions even when they had to reason over potentially competing values. This finding points to potential for developing tools that help users navigate and balance these conflicts. Future work should investigate how people reason about conflicts in value objectives, how we might better communicate these trade-offs, and how to support users in deciding on their ideal value priorities. Furthermore, designing interfaces that clarify input-output mappings --- e.g., through examples --- may enable more meaningful control. One promising direction is to support structured forms of expressivity, such as value clusters (groups of values often sought together) or user-defined combinations using logical operators (e.g., and, or), which can provide finer-grained control and make users' intentions more explicit.

Finally, while Schwartz's theory treats values as relatively stable principles, he also reports that people's values might shift over time in response to life events, changing social contexts, or evolving media habits. For this reason, value-aligned feed interfaces should support revisiting and adjusting value settings when users choose to do so, much like existing controls for interests or personalization. This explicit control maintains user autonomy without requiring continuous monitoring or inference of values. At the same time, platforms already support preference refinement through lightweight, optional feedback mechanisms (e.g., ``show me more of this'' and ``I liked this''). Value-aligned feeds could incorporate analogous interactions, such as letting users indicate when a post ``aligns with my values'', which could be a low-effort way for users to refine their values over time without rearticulating their entire profile.

\subsection{Incorporating Values into Feed Ranking Algorithms}
\label{subsec:ranking}
The profitability of engagement-based ranking for social media companies suggests that value-based ranking, if adopted, would be incorporated alongside engagement metrics in the content curation process. Our work applied value-based reranking at the last stage (i.e., on content that was already filtered for engagement), but real-world systems may incorporate values at earlier stages of candidate generation and relevance estimation. Because the distribution of values in posted content is uneven and further filtered by connection networks and engagement dynamics, introducing value-based ranking earlier in the pipeline can more powerfully reshape what becomes visible. If industry practitioners adopt our approach, they should evaluate the impacts of different insertion points on users, not only on the status quo metrics of user engagement and retention, but also on longer term metrics such as user well-being and attitude. However, in practical terms, we acknowledge that platforms often allow engagement to dominate any ultimate decision-making on what gets launched 
~\cite{christin2024internal}.

Another consideration is how heavily end-user values ought to shape the feed --- a question that connects to broader debates about filter bubbles on social media that arise through the dynamics of selective exposure and personalization~\cite{bozdag2015breaking, kitchens2020understanding}. Users may end up only seeing posts that reinforce their views, limiting their exposure to other values. One way to mitigate this risk is to draw on social choice or other democratic methods to place bounds on how much users can value-shape their feeds. For example, an algorithm might enforce that people can only influence a value's prevalence to within some percentile of the overall population distribution, making it difficult to completely mute competing values. Another approach is to implement bridging algorithms~\cite{tornberg2023simulating} that seek out content that is counter-valued but still broadly appreciated.

A third consideration concerns the ranking function itself. In our implementation, value-based ranking relied on linear weights over content-user value scores. While this choice provides transparency and aligns with standard industry practice of using linear objective functions in ranking, it constrains how complex interactions among values can affect ranking outcomes. More expressive approaches, such as non-linear aggregation or contextualized decision rules, may capture richer value interactions, though they introduce interpretability and governance challenges that merit further study.

\subsection{Choosing Value Systems to Model}
\label{sec:democratized}

While Schwartz's theory provides a research-grounded vocabulary for discussing values, individuals differ in which values a piece of content evokes for them, in part due to their backgrounds, experiences, or values that they personally prioritize among other factors~\cite{haidt2004intuitive}. The labeling scheme in our approach involved providing the list of Basic Human Values, their definitions, and examples of tweets coded by the researchers, which were then used for in-context learning by a large language model. In general, our results suggest that this approach worked: participants recognized the value-aligned feeds that they created. However, variation in perceived value expressions likely contributed to the instances in which users did not recognize their value-aligned feeds. Future work should investigate the factors driving the difference in perceived values in content and build classifiers customized for different groups of users that capture their unique experiences and perceptions. This approach could ensure that value-aligned feeds are more relevant and resonant for various user clusters.

These perceptual differences also raise a broader question of which value system should a feed use as its underlying representational space. Although we treat Schwartz’s theory of Basic Human Values as a practical foundation, it is not the only valid value system. We adopt Schwartz because it is a \textit{complete} value system, in the sense that it seeks to both articulate a full design space and also describe to what extent these values are conflicting or complementary. Additionally, its values are few enough in number that they can be integrated into user interface controls, it is widely studied and replicated, and it provides the value similarities and differences inherent to a complete value system. Other value systems fitting these criteria could replace Schwartz in our approach. While we evaluated our approach in the context of Schwartz’s values, we believe that the stages that it comprises --- from operationalizing values in tweets, to scaling value annotations, to ranking content based on values --- can generalize to values beyond those in Schwartz’s system.

Nonetheless, relying on a predefined value system introduces practical limitations. Not every value in Schwartz's theory is immediately understandable to a lay user. This challenge led us to reword several of the value names in our interfaces in consultation with a cultural psychologist, to aid usability. In addition, even a comprehensive value system may fail to capture values that are culturally specific, nuanced, or especially salient in the context of social media. Therefore, there is an opportunity for extending our work to other values or value systems, and perhaps even those that individuals or communities have identified as important through their own experiences~\cite{weld2024making, weld2022makes}, rather than constraining them to a predefined top-down set of values. Future work should investigate offering users a compilation of these various values and interfaces for capturing them, allowing for a more tailored content curation. These questions are, in fact, larger than social media. With growing concerns about whose values large language models encode and output~\cite{santurkar2023whose,durmus2023towards}, these questions are also highly relevant to the AI alignment community.

\section{Limitations}
We highlight limitations that can also inform future work in this area. First, in our method, we only label values present in the content of the tweet including its text, images, and embedded links. This approach may overlook values that may be evoked from other factors such as the author of the content. Having additional information, such as who is posting the tweet, can modulate how the values in the content are interpreted. Echoing discussion in Sec.~\ref{sec:democratized}, future work that explores how we might personalize value labels and incorporate contextual knowledge into the value labeling process can improve the performance of our ranking method.

A concern with any method that uses large language models is the potential biases latent in the model outputs. Prior work~\cite{benkler2023assessing, tao2024cultural} has found that popular models are more likely to reflect values that are found in Western countries. It is possible that our LLM labels overrepresent values that are more prioritized in Western culture. However, since there are not objective ground-truth labels for values, it is difficult to identify if such bias is present. Nonetheless, promising work~\cite{feng2024modular, sorensen2024value} has proposed methods for incorporating more pluralistic values into models that can be adopted to mitigate potential biases.

In our work, we use recognizability of the resulting feed as the primary evaluation metric for value alignment. We acknowledge that recognizability does not capture all possible forms of alignment inside the model. However, it is the most direct and interpretable measure of whether the system's input–output mapping remains meaningful from the user's perspective.

Some of the limitations of our work relate to the setup of our studies. First, in our controlled experiments, users interacted with the system in a single session. As a result, we cannot predict how long-term use of value-based feed ranking would affect user behavior. A longitudinal study is needed to explore whether and when users would choose to consume value-based feeds over engagement-based ones.
A longitudinal study can also uncover how sustained exposure to value-aligned feeds might influence engagement patterns and any downstream effects. For instance, if users engage more with value-aligned content that is promoted to them, would the engagement-based algorithm then learn to prioritize content with similar values?

Second, when ranking feeds, the available content on the user's feed limited how well we could match their specified value weights. For example, if a user prioritized ``Achievement'' but there were no tweets containing the value, we were not able to amplify it in the resulting feed. This challenge was primarily an artifact of the context we were operating in --- the inventory of tweets at our disposal for each task was ranged from 5 to 9 batches (with each batch consisting of approximately 30 tweets) collected from the user’s algorithmically curated feed. Nevertheless, even given this constraint, our approach yielded value-aligned feeds that users could overall recognize. If social media companies adopt this approach, they can conceivably better satisfy users' value-driven preferences because they have access to the entire inventory of in-network and out-of-network content for a user. 

Finally, we only recruited participants based in the United States. Although Schwartz's Basic Human Values are considered universal across cultures, we cannot predict how our end-to-end ranking method would impact non-U.S. users, as cultural differences can impact what values users prioritize or how they respond to the ranking approach. An avenue of future work is comparing differences in value selections across cultures.

\section{Conclusion}
Social media feeds are inherently value-laden with downstream influences that extend beyond the digital world. While values are currently passively encoded as a byproduct of engagement optimization, we should consider designing ranking algorithms that directly take these values into account. But, if taking this approach, what values should we embedded in these feeds? In this work, we introduce a method for incorporating a comprehensive set of human values with trade-offs, drawn from cultural psychology, into feed ranking algorithms. Since the value constructs are broad by-design, covering the fundamental motivations underlying human behavior, our method generalizes to a diverse range of content topics and users. Through two studies, we validate that this method can indeed embed basic values into algorithms in a manner that is interpretable to users. Furthermore, users can specify the values they want to see on social media and directly reconfigure their feeds to reflect these ideals. We then analyze the values that users prioritize when given value controls over their feeds.

\begin{acks}
We thank members of the Stanford HCI Group and Social Media AI Reading Group for their helpful feedback. This work was sponsored by the Hoffman-Yee Research Grants at the Stanford Institute for Human-Centered
Artificial Intelligence (HAI) and NSF 2403435. Dora Zhao is supported in part by the Brown Institute for Media Innovation and the Paul \& Daisy Soros Fellowship for New Americans. Tiziano Piccardi is supported by Swiss National Science Foundation (Grant P500PT-206953). Zachary Robertson is supported by a Stanford School of Engineering fellowship. Sanmi Koyejo acknowledges support from NSF 2046795 and 2205329, the MacArthur Foundation, Stanford HAI, and Google Inc.
\end{acks}

\bibliographystyle{ACM-Reference-Format}
\bibliography{bibliography}


\newpage
\appendix

\section{Labeling Values in Social Media Content}
We detail the process for labeling values in X/Twitter posts using a large language model. Then, we provide examples of tweets with their corresponding value labels.
\subsection{Method}
\label{sec:labeling_method}
The values are labeled in each tweet on a scale from 0 to 6 using \texttt{GPT-4o} with a temperature of 1. As shown in the prompt below, we provide instructions on how to code values that are based in the language used in the PVQ. The definitions of the tweet, as denoted by \texttt{\$\{conceptDefinitionStr\}} are shown in Table.~\ref{tab:value_names} -- we concatenate the definition with the displayed name. In the prompt, we also include two tweets that were manually coded by the researchers as few-shot examples. 

Our prompting method accommodates different tweet formats. For general posts or reposts, we supply the text body at the end of the prompt. Since \texttt{GPT-4o} is able to process images, we include the URL of the image(s) included in the tweet directly in our prompt. For quotations, we append the quoted tweet to the main body with ``QUOTED'' preceding it. We also include the title and the description of embedded links in the tweet in the string and denote them with ``LINK TITLE'' and ``LINK DESCRIPTION'' respectively. For replies, we average the values over all tweets in the conversation. 
\begin{quote}
    \footnotesize
Consider the following set of concepts, listed as \verb|!<CONCEPT>! : !<DEFINITION>!|:

\verb|${conceptDefinitionsStr}|

For each concept, rate whether the tweet reflects the concept on a Likert scale with options [0, 1, 2, 3, 4, 5, 6], where:

\begin{itemize}
  \item 0 = This post does not reflect this concept at all
  \item 1 = This post reflects this concept a little bit
  \item 6 = This post reflects this concept strongly
\end{itemize}

A concept can be reflected if the tweet supports the concept.

\textbf{Output:} One JSON dictionary of the following format.  
Rating is a dictionary of key value pairs, with each key being a concept and each value your rating for that concept:

{ "Rating": {"Concept": !<RATING>!} }

\textbf{Example 1:}

Tweet:  
I've always believed in the power of research to save lives and ensure Americans get the care they need. Starting today,  
the first-ever White House Initiative on Women's Health Research will work towards that goal, changing how we approach  
and fund women's health research.

\{"Rating": \{"Reputation": 0, "Power": 0, "Wealth": 0, "Achievement": 0,
"Pleasure": 0, "Independent thoughts": 0, "Independent actions": 0,
"Stimulation": 0, "Personal security": 0, "Societal security": 0,
"Tradition": 0, "Lawfulness": 0, "Respect": 0, "Humility": 0,
"Responsibility": 5, "Caring": 4, "Equality": 5,
"Connection to nature": 0, "Tolerance": 3\}\}

\textbf{Example 2:}

Tweet:  
This little kid was carrying a white flag, and now he's dead. This guy was also carrying a white flag, and he's been  
shot. I'm here filming for you, and I'm in a lot of danger as well. People holding white flags are trying to come out, and  
are scared from the snipers. If the claims of civilians with white flags getting INTENTIONALLY targeted is true, this  
would be a WAR CRIME.

\{"Rating": \{"Reputation": 0, "Power": 0, "Wealth": 0, "Achievement": 0,
"Pleasure": 0, "Independent thoughts": 0, "Independent actions": 0,
"Stimulation": 0, "Personal security": 0, "Societal security": 0,
"Tradition": 0, "Lawfulness": 0, "Respect": 0, "Humility": 0,
"Responsibility": 0, "Caring": 6, "Equality": 5,
"Connection to nature": 0, "Tolerance": 3\}\}

Tweet:  
\end{quote}

\begin{figure*}
    \centering
    \includegraphics[width=\linewidth]{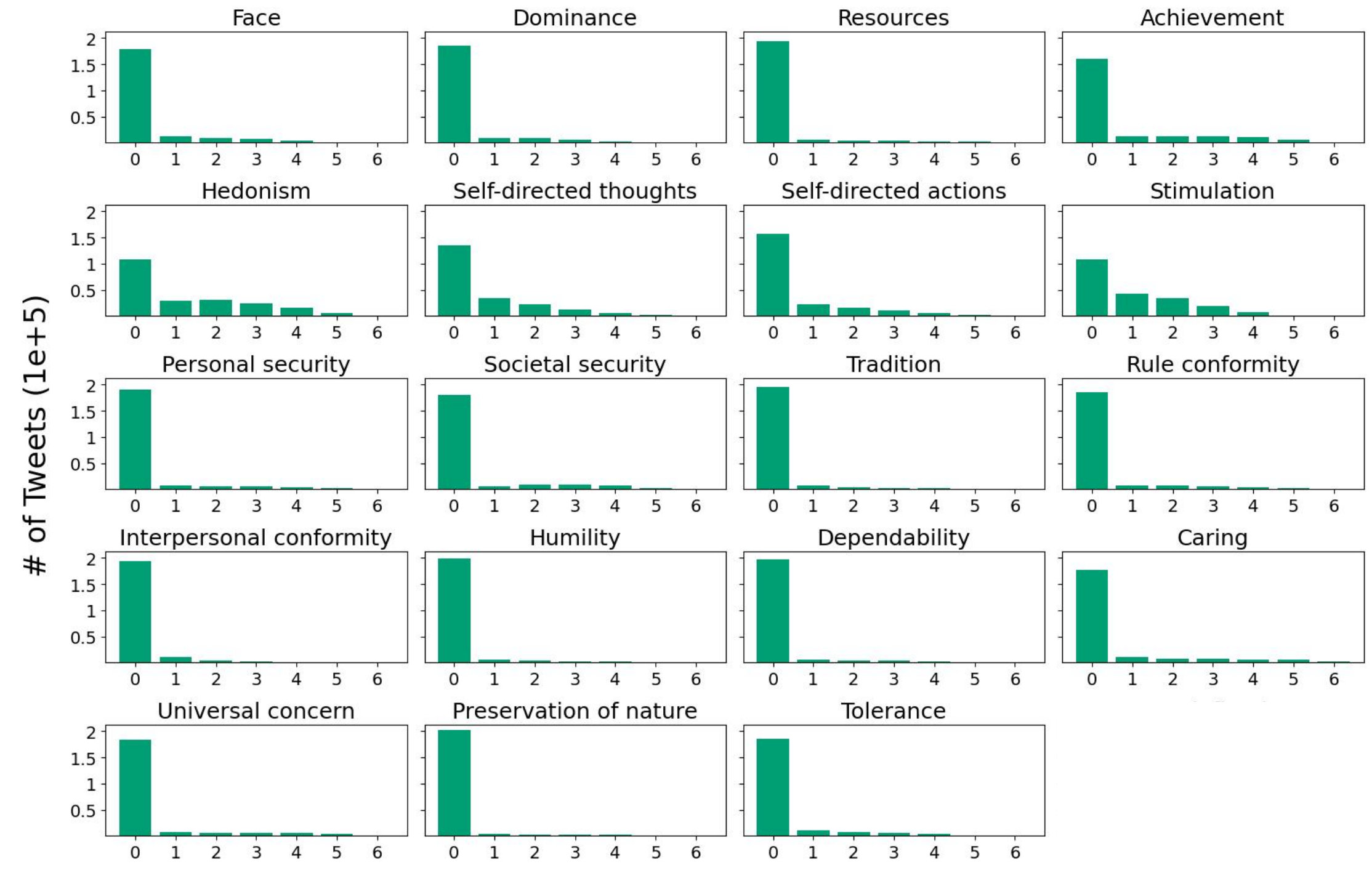}
    \caption{Values are sparsely distributed across tweets (i.e., for a given value, the majority of tweets do not contain this value). We visualize the distribution of labeled values for all 212,663 tweets collected in Study 2. The most prevalent values in tweets are ``Stimulation'' and ``Hedonism.''}
    \label{fig:facet}
    \Description[Distribution of value labels across tweets]{
   The figure shows histograms for all nineteen values, indicating how frequently each value appears across 212,663 tweets. Most tweets contain no explicit value labels, with counts heavily concentrated at zero across all categories, demonstrating that values are sparsely expressed. Among the labeled values, Stimulation and Hedonism appear most frequently relative to others.}
\end{figure*}
\subsection{Labeled Values}
We provide examples of tweets with their corresponding value labels (see Table~\ref{tab:label_ex}. In addition, we provide summary statistics on the values labeled across the 212,663 tweets collected from Study 2 (see Fig.~\ref{fig:facet}). Values are sparsely distributed, meaning that for a majority of tweets the given value is equal to 0. However, 94.4\% of tweets are value-laden (i.e., have at least one value with a label greater than 0). 

\begin{table*}[h]
    \small
    \centering
    \begin{tabular}{p{2.25in}p{3.25in}}
    \toprule
    \textbf{Tweet} &\textbf{Label}  \\ \midrule
    ``I got to be an Amazon VP because a series of managers fought to promote me and pay me more. This was no accident. I used `The Magic Loop' which I invented with my first manager. Here is that story and how you can do it yourself today.'' & \{``Self-directed thoughts'': 4, Self-directed actions'': 5, ``Stimulation'': 2, ``Hedonism'': 1, ``Achievement'': 6, ``Dominance'': 3, ``Resources'': 0, ``Face'':2, ``Personal security'': 0, ``Societal security'': 0, ``Tradition'': 0, ``Rule conformity'': 0, ``Interpersonal conformity'': 0, ``Humility'': 2, ``Dependability'': 0, ``Caring'': 0,
    ``Universal concern'': 0, ``Preservation of nature'': 0, \newline``Tolerance'': 0\} \\\\
    In retrospect, all the `double-jointed' `12,000 calorie diet' and `6’8 wingspan' claims were blatant misdirection to distract from the fact that Phelps was almost certainly doped to the gills QUOTED: @Srirachachau Take away these medals that only happened because of biological advantages & \{``Self-directed thoughts'': 0, ``Self-directed actions'': 0, ``Stimulation'': 0, ``Hedonism'': 0, ``Achievement'': 1, ``Dominance'': 0, ``Resources'': 0, ``Face'': 2, ``Personal security'': 0, ``Societal security'': 0, ``Tradition'': 0, ``Rule conformity'': 0, ``Interpersonal conformity'': 2, ``Humility'': 0, ``Dependability'': 0, ``Caring'': 0, ``Universal concern'': 1, ``Preservation of nature'': 0,\newline``Tolerance'': 0\} \\\\
    Just did a walkthrough of the Tesla supercompute cluster at Giga Texas (aka Cortex). This will be ~100k H100/H200 with massive storage for video training of FSD \&; Optimus. Great work by the Tesla team! & \{``Self-directed thoughts'': 0, ``Self-directed actions'': 0, ``Stimulation'': 2, ``Hedonism'': 0, ``Achievement'': 5, ``Dominance'': 0, ``Resources'': 0, ``Face'': 0, ``Personal security'': 0, ``Societal security'': 0, ``Tradition'': 0, ``Rule conformity'': 0, ``Interpersonal conformity'': 0, ``Humility'': 0, ``Dependability'': 0, ``Caring'': 0, ``Universal concern'': 0, ``Preservation of nature'': 0,\newline``Tolerance'': 0\} \\\\
    Hearing a number of high-level Democrats reached out to the Kamala camp last night and told them that Shapiro would be a `substantially precarious' choice at this point & \{``Self-directed thoughts'': 0, ``Self-directed actions'': 0, ``Stimulation'': 0, ``Hedonism'': 0, ``Achievement'': 0, ``Dominance'': 0, ``Resources'': 0, ``Face'': 1, ``Personal security'': 0, ``Societal security'': 0, ``Tradition'': 0, ``Rule conformity'': 0, ``Interpersonal conformity'': 0, ``Humility'': 0, ``Dependability'': 0, ``Caring'': 0, ``Universal concern'': 0, ``Preservation of nature'': 0,\newline``Tolerance'': 0\} \\\\
    \#BREAKINGNEWS It's now confirmed NABJ and Donald Trump had a disagreement about fact checking him live. Trump refused to go on stage until they agreed to his demands. NABJ president Ken Lemon wrote an actual statement of why Trump wouldn't be appearing and was prepared to present it when Trump suddenly walked on stage.& \{``Self-directed thoughts'': 1, ``Self-directed actions'': 2, ``Stimulation'': 2, ``Hedonism'': 0, ``Achievement'': 0, ``Dominance'': 3, ``Resources'': 0, ``Face'': 2, ``Personal security'': 0, ``Societal security'': 0, ``Tradition'': 0, ``Rule conformity'': 1, ``Interpersonal conformity'': 1, ``Humility'': 0, ``Dependability'': 1, ``Caring'': 0, ``Universal concern'': 0, ``Preservation of nature'': 0,\newline``Tolerance'': 1\} \\
    \bottomrule
    \end{tabular}
    \caption{We provide five examples of tweets and their value labels, generated from the labeling method described in Sec.~\ref{sec:labeling_method}}
    \label{tab:label_ex}
\end{table*}

\section{Pilot Studies}
\label{sec:pilot}
We conducted three pilot studies: one for Study 1 and two for Study 2. For our pilot of Study 1, we recruited 30 participants from Prolific. To qualify for the study, participants must reside in the United States, be over 18 years old, have a Twitter account, and have over 250 submissions on Prolific with a $\geq 95\%$ acceptance rate. Participants received \$5.00 for completing the task (prorated from \$15.00 per hour). From the pilot, the mean recognizability rate was $75.0\%$.

For Study 2, we first ran a pilot where participants could change up to six sliders corresponding to different values. We recruited 150 participants from Prolific using the same screening criteria as for Study 1. Participants received \$5.75 for taking part in the study. From this pilot, we observed a large drop-off for participants who change more than one slider although the decrease in recognizability plateaus. Thus, for cost-saving measures, in the remaining pilots and in our final study, we allow participants to change up to five sliders or all nineteen. In the second pilot, we again recruited participants from Prolific using the same screening criteria. In total, we had 35 participants take part in the pilot; they received \$5 in compensation (prorated from \$15 per hour). In the study, participants are shown all 19 sliders but can adjust 1, 2, 3, 4, 5, or all 19 depending on which condition they have been randomly assigned. From our pilot results, we observe a mean recognizability of $70.5\%$ across conditions.

\section{Think-Aloud Study}
To better understand how participants select value sliders, we conduct a think-aloud study where participants interact with the interface from Study 2. To note, these participants are distinct from those that took part in Studies 1 and 2.

\subsection{Method}
We provide details on how the think-aloud study was conducted and the analysis method. This study was approved by Stanford University's Institutional Review Board.

\subsubsection{Study Design}
In the think-aloud study, participants completed the same protocol as described in Study 2 (Sec.~\ref{sec:study2}) and were given access to all 19 sliders (i.e., the Full Sliders condition). We instructed the participants to think aloud as they adjusted their value sliders and reason about which feed they believed was value-ranked. Each session lasted approximately 35–60 minutes, and participants received a \$20 gift card via Tremendous as compensation. The think-aloud studies were video- and audio-recorded with participants' consent.

\subsubsection{Participants} In total, we recruited thirteen participants via university mailing lists. We include information about participant demographics in Table~\ref{tab:think_aloud}. The median age of participants is $23.5$ years old. Nine of the participants identified as female, three as male, and one as non-binary. All but one of our participants use Twitter at least a few times a week.

\begin{table*}[hb]
    \centering
    \begin{tabular}{lrrr}
    \toprule 
    Participant &  Twitter Usage  & Gender & Race\\
    \midrule 
    P1 &  A few times a week & Female & Asian\\
    P2 & Every day & Male & Asian\\ 
    P3 & A few times a month & Female & Asian\\
    P4 & A few times a week & Male & Asian\\
    P5 & Every day & Female & White, Hispanic\\
    P6 & Every day & Female & White\\ 
    P7 & A few times a week & Female & Black \\
    P8 & A few times a week & Female & Asian\\
    P9 & Every day & Non-binary & Asian\\
    P10 & A few times a week & Male & Asian\\ 
    P11 & A few times a week & Female & White, Hispanic\\
    P12 & Every day & Female & Asian \\
    P13 & A few times a week & Female & White, Asian \\
    \bottomrule
    \end{tabular}
    \caption{Summary of demographic background and Twitter / X usage frequency for participants in our think-aloud study.}
    \label{tab:think_aloud}
\end{table*}
\subsubsection{Analysis}
We analyzed the think-aloud sessions using an inductive, collaborative approach focused on identifying common patterns in how participants interacted with the slider interface and reasoned about value selections. We used Dovetail\footnote{https://dovetail.com/} to transcribe the recordings from the think-aloud study. Then, two researchers independently read the transcripts, noting recurring behaviors and decision strategies. Then, we discussed these observations and iteratively developed a set of themes characterizing how participants interpreted the values, adjusted their sliders, and attempted to infer which feed was value-ranked. Using these collaboratively generated categories, both researchers revisited the transcripts and refined the themes through discussion. The resulting themes reflect consistent patterns in how participants understood the values, prioritized or deprioritized them, and navigated tensions between competing values.

\subsection{Results}
\noindentparagraph{Topical interests and platform norms shaped value selections.} Beyond personal values, we also find that participants often mapped the \emph{content} they wanted to see to the underlying values. Many participants approached the task with specific topics in mind (e.g., popular culture, AI papers) and used the sliders as a way to promote or demote those types of content in their feeds. For example, P3 was interested in ``\emph{fairness research in CS,}'', upranking values related to ``responsibility'' and ``equality.'' Similarly, P4 described their strategy as ``\emph{how do I tune those parameters so that I get AI-related tweets?}'' leading them to uprank ``Achievement''.

Participants also noted the discourse around a particular value varies across platforms. Even when a slider corresponded to a value they personally resonated with and wanted to prioritize, some participants hesitated because they anticipated that posts reflecting that value would be misunderstood or derailed in the platform's discourse. As P7 explains, even though they personally valued ``Preservation of Nature'', they downranked the slider since they felt ``\emph{There's some topics that you can't talk about on Twitter because people are just going to amplify the worst part of it... I guess I do want to see it on my feed if people were able to actually understand it.}''

\noindentparagraph{Values are a bridge across topics.}
Participants found values to be a useful construct for re-ranking feeds because they can act as a common denominator underlying multiple topics of interest. For example, P4 initially upranked ``Achievement'' to surface more AI-related content but discovered that this also brought up posts about a motivational speaker they liked. They reflected: ``\emph{The sliders are like the factor analysis. Those are the factors that can explain a lot of variation... it could also give me content that I wasn’t thinking about, but it’s usually highly correlated with the content that I like.}'' In addition to bridging topics, values can also help connect across perspectives. For instance, P11 found the sliders helpful in surfacing content from opposing points of view that still might resonate with them, as opposed to the ``rage-bait'' they think the engagement feed currently prioritizes: ``\emph{it would be nice to see the other side without it being presented to me in such a monstrous way because often times what [the engagement feed] is showing to me is like I'm seeing the worst of it.}'' However, using values as this broad basis for reasoning over content also introduced challenges. Several participants reported difficulty forming a clear mental model of how slider adjustments would affect their feeds. As P1 explained: ``\emph{For some of the constructs in my head it was unclear as to how it’s going to be translated to re-ranking a particular feed.}''

\noindentparagraph{Utility depends on the purpose of the account.}
The perceived usefulness of value-based controls varied by participants’ goals and platforms. Several who used Twitter mainly for professional purposes found the controls less relevant, while others, like P4 and P6, preferred using them on personal accounts or when browsing for exploration. For participants, value-based controls were especially appealing to help shape engagement-driven feeds, such as the ``For You'' feed on Twitter or Threads, as opposed to their ``Following'' feed. As P8 explained that sometimes it felt as if ``\emph{[the algorithm knows] me too well}''. When they wanted to shift the content they were seeing, they felt as if they had no control  to do so ``\emph{because it's based on my algorithm, like how I interacted with the previous post and there's no way to change that.}'' Finally, participants’ interest depended on whether they frequently encountered value-laden content on the platform. P7, for example, wanted value controls on Instagram, where they saw posts about ``\emph{news and helping people,}'' compared to Twitter, which they used for ``\emph{unserious purposes.}''

\begin{figure*}
    \centering
    \includegraphics[width=\textwidth]{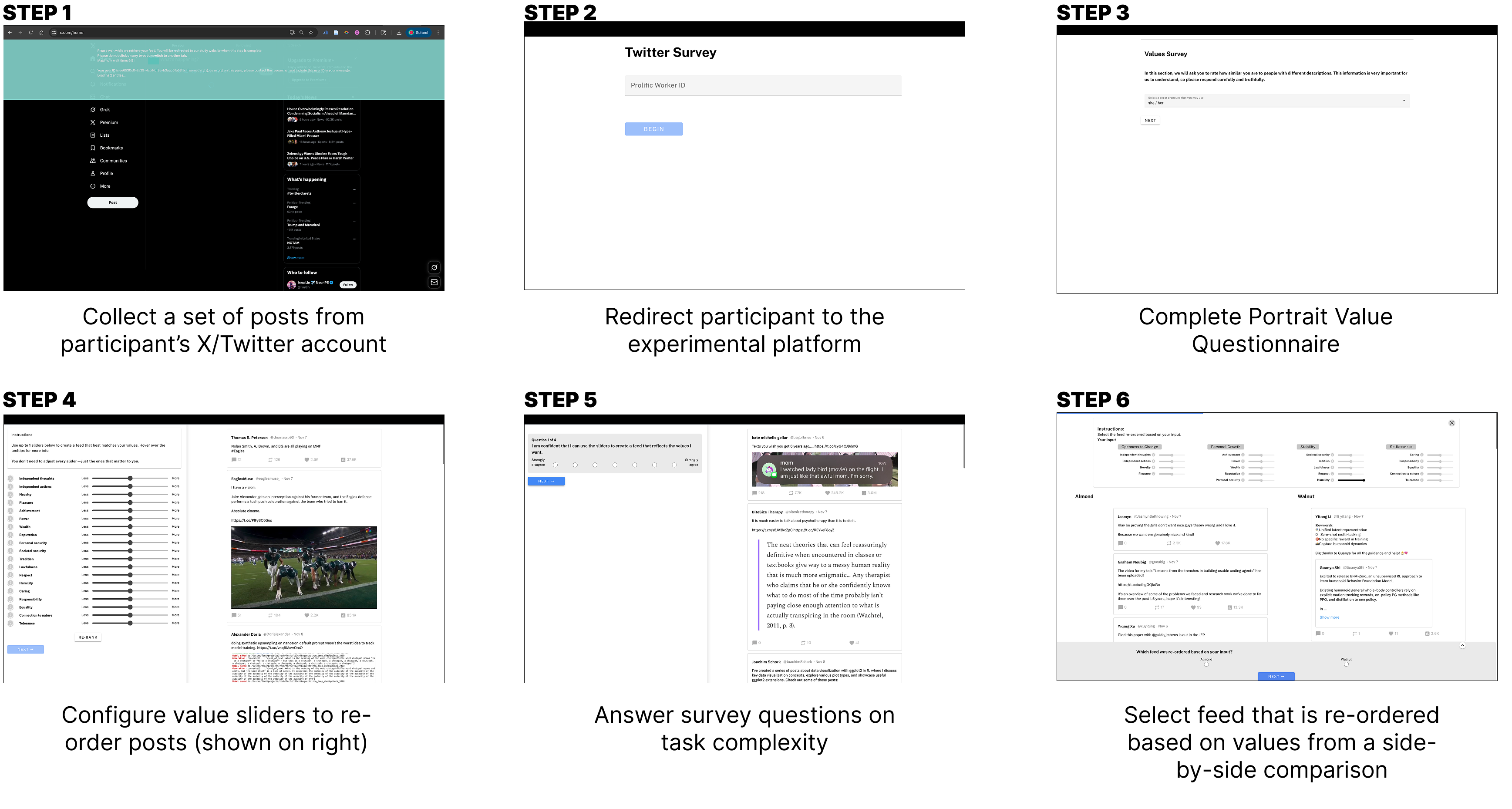}
    \caption{Overview of the study design flow with screenshots of the user interfaces.}
    \label{fig:survey_flow}
    \Description[Overview of the six-step study procedure]{
   The figure illustrates the full study workflow. First, participants provide access to their X/Twitter posts, which are collected for analysis. Second, they are redirected to the experimental survey platform. Third, participants complete a Portrait Value Questionnaire to assess personal values. Fourth, they adjust value sliders that dynamically reorder their social media posts based on selected values. Fifth, participants answer follow-up survey questions about task complexity and experience. Finally, they choose between different feed versions in a side-by-side comparison, selecting the one they prefer based on value alignment.
    }
\end{figure*}
\section{Additional Study Details}
First, we provide an overview of our study design in Fig.~\ref{fig:survey_flow}. Second, we provide more information on the demographics of participants in both Study 1 and Study 2 as follows. In Study 1, our participants have a median age of $37.0$ years old. The sample is slightly skewed towards participants who identify as male ($52.5\%$). $45.4\%$ of participants identify as female, and $2.1\%$ as non-binary. Since we stratified our sampling of political leanings into two categories --- Democrat and Republican or Independent, our participants are approximately balanced: $37.6\%$ are Democrats, $29.1\%$ are Independents, and $30.5\%$ are Republicans. With respect to race, $68.8\%$ identified as White, $13.5\%$ as Black, $5.7\%$ as Asian or Pacific Islander, $1.4\%$ as Indigenous  / Native American, and $10.6\%$ as Multi-racial. Our participants are frequent users of Twitter with $62.7\%$ visiting the platform at least once a day.

The demographics of participants in Study 2 closely mirror those of Study 1. The median age of participants was $34.0$ years old. 45.0\% of participants identified as female, $52.2\%$ as male, and $2.8\%$ as non-binary. 52.2\% identified as Democrat, 12.4\% as Republican, 32.1\%
as Independent, and 3.2\% as Other. With respect to race, $59.0\%$ identified as White, $18.5\%$ as Black, $8.8\%$ as Asian, and $7.4\%$ as Multi-racial. Similar to Study 1, over half ($51.0\%$) of participants visited Twitter at least once a day. 

\subsection{Validating Labels}
\label{sec:app_mae}
\begin{table*}[]
    \centering
    \begin{tabular}{lrrr}
    \toprule
        Value &  Human-Consensus MAE & LLM-Consensus MAE \\
    \midrule
    Self-directed thoughts & $1.06\pm1.10$ & $1.06\pm1.14$\\
Self-directed actions & $1.08\pm1.12$ & $1.08\pm1.18$\\
Stimulation & $0.93\pm1.04$ & $1.05\pm1.12$\\
Hedonism & $1.23\pm1.07$ & $1.02\pm1.10$\\
Achievement & $0.93\pm1.07$ & $0.98\pm1.24$\\
Dominance & $0.63\pm0.83$ & $0.60\pm0.90$\\
Resources & $0.69\pm0.90$ & $0.53\pm0.82$\\
Face & $1.17\pm1.07$ & $0.99\pm1.07$\\
Personal security & $0.46\pm0.82$ & $0.63\pm1.11$\\
Societal security & $1.10\pm1.04$ & $0.96\pm1.07$\\
Tradition & $1.05\pm1.01$ & $0.80\pm0.91$\\
Rule conformity & $1.05\pm1.06$ & $0.83\pm1.03$\\
Interpersonal conformity & $0.94\pm0.96$ & $0.90\pm1.11$\\
Humility & $1.53\pm0.90$ & $1.41\pm1.07$\\
Dependability & $1.49\pm1.03$ & $1.26\pm1.14$\\
Caring & $1.62\pm1.06$ & $1.31\pm1.19$\\
Universal concern & $1.38\pm1.08$ & $1.16\pm1.18$\\
Preservation of nature & $0.76\pm0.91$ & $0.59\pm0.90$\\
Tolerance & $1.20\pm1.00$ & $0.95\pm0.98$\\
\midrule
Overall & $1.07\pm1.05$ & $0.95\pm1.10$\\
\bottomrule
    \end{tabular}
    \caption{Language models produce labels that align more closely with the consensus label compared to human annotators. We validate our LLM labeler on a dataset of 4,503 Twitter/X posts that had been labeled by three or more annotators. For each value, we computed the mean absolute error between any individual annotator and the remaining annotators (Human-Consensus MAE) and the difference between the LLM and the average annotators.}
    \label{tab:mae}
\end{table*}

We conduct an intrinsic evaluation of our LLM labeler on a dataset of 4,503 Twitter/X posts from \citet{epstein2025measuring} that have been labeled by at least four annotators. The number of labels per post range from four to ten. Prior works have used human annotators to label expressions of Schwartz's values in text content, including in written arguments~\cite{kiesel2022identifying}, social scenarios~\cite{qiu2022valuenet}, and social media content~\cite{epstein2025measuring} Following these works, we filter to a set of posts that have been labeled by multiple human annotators to ensure higher confidence in the ground truth annotations, selecting a conservative threshold of posts with at least four annotators. We report the disaggregated errors between any individual annotator and the remaining annotators as well as the difference between the LLM and the average annotations in Table~\ref{tab:mae}.

In addition, we also report the LLM-Consensus RMSE, or root mean-squared error, as well as the LLM-Consensus RMSE. Compared to MAE, RMSE is more sensitive to large errors, making it useful for identifying cases where the LLM's predictions deviate substantially from the consensus. As reported in Table~\ref{tab:rmse}, the LLM-Consensus RMSE is $1.45$ whereas the Human-Consensus RMSE is $1.74$, indicating that the LLM performs as well, if not better, than the average annotator at predicting the consensus. Coupled with the LLM-Consensus MAE results, this indicates that our LLM labeler not only agrees closely with the consensus on average, but also avoids extreme mispredictions.

\begin{table*}[]
    \centering
    \begin{tabular}{lrrr}
    \toprule
        Value &  Human-Consensus RMSE & LLM-Consensus RMSE \\
    \midrule
Self-directed thoughts & $1.76$ & $1.55$\\
Self-directed actions & $1.80$ & $1.60$\\
Stimulation & $1.62$ & $1.53$\\
Hedonism & $1.87$ & $1.50$\\
Achievement & $1.65$ & $1.58$\\
Dominance & $1.28$ & $1.09$\\
Resources & $1.36$ & $0.98$\\
Face & $1.84$ & $1.46$\\
Personal security & $1.15$ & $1.27$\\
Societal security & $1.75$ & $1.44$\\
Tradition & $1.70$ & $1.21$\\
Rule conformity & $1.72$ & $1.32$\\
Interpersonal conformity & $1.58$ & $1.43$\\
Humility & $2.02$ & $1.77$\\
Dependability & $2.06$ & $1.70$\\
Caring & $2.19$ & $1.77$\\
Universal concern & $1.99$ & $1.65$\\
Preservation of nature & $1.41$ & $1.08$\\
Tolerance & $1.81$ & $1.36$\\
\midrule
Overall & $1.74$ & $1.45$\\
\bottomrule
    \end{tabular}
    \caption{For each value, we computed the root mean-squared error between any individual annotator and the remaining annotators (Human-Consensus RMSE) and the difference between the LLM and the average annotator (LLM-Consensus RMSE).}
    \label{tab:rmse}
\end{table*}

\subsection{Robustness Analysis}
\subsubsection{Recognizability Measure} 
\label{sec:recog_robustness}
In Studies 1 and 2, we measure recognizability by asking the participant to select the feed they believe is more aligned to a single value or their selected values respectively. They are shown two side-by-side feeds: one feed is the value-aligned feed and the other is the engagement feed. As a robustness check, we replicated results from Study 2 when participants are shown one slider and all nineteen sliders in which participants are shown either their value-aligned feed or a random reshuffling of posts, both from the same batches of tweets. We recruited 80 participants in total; of these, 11 passed the attention checks, leaving us with 69 participants --- 32 are allowed to adjust one slider and 37 are allowed to adjust all nineteen.

In line with results from Study 2, participants in the one slider condition had a mean recognizability of $71.1\% \pm 1.8$. For participants in the full sliders condition, mean recognizability was higher than in the original study at $69.4\% \pm 1.7$ compared to $61.0\% \pm 1.3$. It is possible that the engagement feed provides a more difficult baseline to compare against relative to a random reshuffling as there may be a small extent of user values embedded in the original ranking.

\subsubsection{Sliders Changed versus Sliders Available} In Study 2, we report a linear mixed-effect model using \texttt{Sliders Changed} as our fixed effect. This predictor corresponds to the number of sliders that the participant actually changed, which may be fewer than the maximum available. As a robustness check, we report the results using the assigned condition (\texttt{Sliders Available}) as a fixed effect. Similar to the results reported in Sec.~\ref{sec:study2}, we observe a weak relationship between \texttt{Sliders Available} and recognizability ($\beta=-0.028$, $p=0.142$). In addition, when we binarize \texttt{Sliders Available}, separating the condition when participants can only change one slider versus when they can change more than one, we observe a significant, negative association with recognizability ($\beta=-0.578$, $p = 0.03$).


\end{document}